\documentclass[a4paper,amsfonts, amssymb, amsmath, reprint, showkeys, nofootinbib, twoside, superscriptaddress]{revtex4-1}
\usepackage[english]{babel}
\usepackage[utf8]{inputenc}
\usepackage[colorinlistoftodos, color=green!40, prependcaption]{todonotes}
\usepackage{amsthm}
\usepackage{mathtools}
\usepackage{physics}
\usepackage{xcolor}
\usepackage{graphicx}
\usepackage[left=20mm,right=20mm,top=35mm,columnsep=15pt]{geometry} % 23 13
\usepackage{adjustbox}
\usepackage{placeins}
\usepackage[T1]{fontenc}
\usepackage{lipsum}
\usepackage{csquotes}
\usepackage{siunitx}
\usepackage{caption}
\usepackage{subcaption}
\usepackage[pdftex, pdftitle={Article}, pdfauthor={Author}, colorlinks=true]{hyperref}
\begin{document}

%%%------------------------------------------------

\title{
% Comparing Simulations of the Quantum Gravity Induced Entanglement of Masses (QGEM) experiment with 2 and 3-qubits\\
% or\\
% New 3-Qubits Quantum Gravity Induced Entanglement of Masses (QGEM) protocol: Quantum simulations in an experimental setup\\
% or\\
% Extention of the Quantum Gravity Induced Entanglement of Masses (QGEM) to tripartite qubits systems.\\
% or\\
Improving resilience of the Quantum Gravity Induced Entanglement of Masses (QGEM) to decoherence using 3 superpositions.}
\author{Martine Schut}
    \email[]{martine.schut@rug.nl}
    \affiliation{Van Swinderen Institute for Particle Physics and Gravity, University of Groningen, 9747AG Groningen, the Netherlands \vspace{1mm}}
    \affiliation{Bernoulli Institute for Mathematics, Computer Science and Artificial Intelligence, University of Groningen, 9747 AG Groningen, the Netherlands \vspace{1mm}}
\author{Jules Tilly}
    \email{jules.tilly.18@ucl.ac.uk}
    \affiliation{Department of Physics and Astronomy, \\ University College London, Gower Street, WC1E 6BT London, United Kingdom \vspace{3mm}}
\author{Ryan J. Marshman}
    \email{r.marshman@uq.edu.au}
    \affiliation{Department of Physics and Astronomy, \\ University College London, Gower Street, WC1E 6BT London, United Kingdom \vspace{3mm}}
\author{Sougato Bose}
    \email{s.bose@ucl.ac.uk}
    \affiliation{Department of Physics and Astronomy, \\ University College London, Gower Street, WC1E 6BT London, United Kingdom \vspace{3mm}}
\author{Anupam  Mazumdar \vspace{1.5mm}}
    \email{anupam.mazumdar@rug.nl}
    \affiliation{Van Swinderen Institute for Particle Physics and Gravity, University of Groningen, 9747AG Groningen, the Netherlands \vspace{1mm}}

\date{\today} 

%%%------------------------------------------------

\begin{abstract}
Recently a protocol called quantum gravity induced entanglement of masses (QGEM) that aims to test the quantum nature of gravity using the entanglement of 2 qubits was proposed. The entanglement can arise only if the force between the two spatially superposed masses is occurring via the exchange of a mediating virtual graviton. In this paper, we examine a possible improvement of the QGEM setup by introducing a third mass with an embedded qubit, so that there are now 3 qubits to witness the gravitationally generated entanglement. We compare the entanglement generation for different experimental setups with 2 and 3 qubits and find that a 3-qubit setup where the superpositions are parallel to each other leads to the highest rate of entanglement generation within $\tau = 5 \, \si{\s}$.
We will show that the 3-qubit setup is more resilient to the higher rate of decoherence.
The entanglement can be detected experimentally for the 2-qubit setup if the decoherence rate $\gamma$ is $\gamma<0.11 \, \si{\hertz}$ compared to $\gamma<0.16 \, \si{\hertz}$ for the 3-qubit setup.
However, the introduction of an extra qubit means that more measurements are required to characterise entanglement in an experiment. We conduct experimental simulations and estimate that the 3-qubit setup would allow detecting the entanglement in the QGEM protocol at a $99.9\%$ certainty with $\order{10^4}-\order{10^5}$ measurements when $\gamma \in [0.1,0.15] \, \si{\hertz}$. Furthermore, we find that the number of needed measurements can be reduced to $\order{10^3}-\order{10^5}$ if the measurement schedule is optimised using joint Pauli basis measurements. %with respect to the measurements of Pauli basis elements jointly.
For $\gamma > 0.06 \, \si{\hertz}$ the 3-qubit setup is favourable compared to the 2-qubit setup in terms of the minimum number of measurements needed to characterise the entanglement. Thus, the proposed setup here provides a promising new avenue for implementing the QGEM experiment.
\end{abstract}

%%%------------------------------------------------
%\keywords{first keyword, second keyword, third keyword}

\maketitle

%%%------------------------------------------------
\section{Introduction} \label{sec:intro}

The quantum or classical nature of gravity has long been a central theme in theoretical physics~\cite{Kiefer:2014sfr}. Unlike other interactions of nature gravity remains the only known interaction whose quantum behaviour has never been observed in a lab. It is believed that the spin-2 massless graviton is the carrier of gravitational interaction, which can be canonically quantised around a Minkowski background \cite{gupta1954gravitation,Feynman:1963ax,DeWitt:1967yk,DeWitt:1967ub}. However, the direct detection of a graviton remains extremely difficult due to the weakness of the gravitational interaction compared to the other fundamental interactions of nature \cite{dyson2014graviton,Baym:2009zu}. Even the direct detection of primordial gravitational waves will not be able to falsify the quantum nature of graviton~\cite{Ashoorioon:2012kh,Martin:2017zxs}. The indirect detection of the quantum aspects of graviton may be feasible in the near future in laboratory-based experiments.

Recently, there has been a proposal to witness the quantum nature of gravity by witnessing the entanglement of masses~\cite{bose2017spin}, where the theoretical protocol was outlined in~\cite{bose2017spin,marshman2020locality}. Simultaneously, there was another paper \cite{Marletto_2017} where the authors proposed to witness the quantum nature of gravity by witnessing entanglement. However, the detailed analysis of the graviton as a mediator, and the feasibility aspects of the experiment including decoherence and the relevant background were first discussed in Ref.~\cite{bose2017spin}.
% alternatively:
% Recently, there has been an idea and an experimental proposal to witness quantum nature of gravity by witnessing the entanglement of masses \cite{bose2017spin} detailing the implementation parameters and decoherence estimates, with the theoretical assumptions and mediation mechanism clarified later in \cite{marshman2020locality}. 
% The "no entanglement generation by Local Operations and Classical Communication (LOCC)" was used as a crucial justification in these papers \cite{bose2017spin,marshman2020locality}, see also \cite{Marletto_2017} where just the idea was presented, but accompanied by a different argument.
The proposal was followed by extensive interest in the research community, suggesting extensions and variations \cite{Belenchia2018,nguyen2019entanglement,pedernales2019motional,marshman2021large,van2020quantum,Toros:2020krn,chevalier2020witnessing,torovs2021relative,christodoulou2019possibility,Christodoulou_2020,margalit2021realization,tilly2021qudits,Carney:2021yfw,howl2020testing,Miki:2020hvg,Matsumura:2020law,qvarfort2020mesoscopic,miao2020quantum,weiss2021large,datta2021signatures,krisnanda2020observable,Haine2021}.
The protocol is based on a \textit{bonafide} \emph{quantum mechanical} gravitational interaction which cannot be replicated by a classical world. 
%The three main requirements are: (1) creation of the spatial quantum superposition of masses, i.e. creating Schr{\"o}dinger cat states, in a laboratory, (2) quantum entanglement, which yields quantum correlation between any two quantum states and has no classical analogue~\cite{1935PCPS...31..555S}, and (3) reliance on the local operation and classical communication (LOCC)~\cite{bennett1996mixed,horodecki2009quantum} principle. 
One of the main experimental requirements is the creation of the spatial quantum superposition of masses, i.e. creating Schr{\"o}dinger cat states, in a laboratory.
For the detection of quantum aspects of gravity we require the generation of quantum entanglement, which yields quantum correlations between any two quantum states and has no classical analogue~\cite{1935PCPS...31..555S}.
We assume that no other quantum interactions arise between the systems.
To prove that the detection of entanglement shows the quantum nature of gravity we rely on the properties of the Local Operation and Classical Communication (LOCC) principle ~\cite{bennett1996mixed,horodecki2009quantum}.
The LOCC principle states that the two quantum states cannot be entangled via a classical channel if they were not entangled to begin with, or entanglement cannot be increased by local operations and classical communication. The classical communication is the critical ingredient which can be put to test when it comes to graviton mediated interaction between the two masses. 
%Following this principle, if entanglement is detected between the qubits used in the experiment, it implies a quantum medium allows them to interact. Provided one eliminates all interaction medium outside of gravity, it means the graviton would give rise to 
If the graviton is quantum, it could mediate the gravitational attraction between the two masses and it would also entangle them, hence giving rise to the quantum gravity induced entanglement of masses (QGEM) proposal \cite{bose2017spin,marshman2020locality}. The QGEM protocol highlighted the requirement of a graviton as a quantum mediator, implying the origin of the gravitational force between masses can be viewed as resulting from the exchange of a virtual graviton. A virtual graviton is not a classical entity, and does not satisfy the classical equations of motion, there are total six
off-shell degrees of freedom, i.e., spin-2 and spin-0 components in the graviton propagator, see~\cite{van1973ghost,biswas2013nonlocal,marshman2020locality}. By witnessing the entanglement between the two masses, and by detecting a correlation between the spins which are embedded in the two test masses, we can ascertain whether the exchange of the virtual graviton is a classical or a quantum entity~\cite{bose2017spin,marshman2020locality}.
%[AM: I have removed it, it does not make sense to me, there are no alternative ways of testing quantum nature of graviton in my opinion. %or alternatives \cite{howl2020testing,Miki:2020hvg,Matsumura:2020law,Haine2021,Rijavec:2020qxd,Bruschi2020,tilly2021qudits,parikh2021quantum} to the experiment.
%\textbf{Because the QGEM protocol is experimentally realizable with current technologies \cite{marshman2021large,margalit2021realization}, it is more feasible on the short term compared to other ways of testing quantum gravity \cite{Haine2021,Bruschi2020,parikh2021quantum}.}

Many possible improvements to the original setup have already been discussed, using a different witness \cite{chevalier2020witnessing}, ways of creating macroscopic quantum superposition \cite{marshman2021large}, including effects due to decoherence \cite{chevalier2020witnessing,Rijavec:2020qxd}, ameliorating Casimir induced entanglement~\cite{van2020quantum}, taming gravity gradient noise and relative acceleration~\cite{Toros:2020krn,torovs2021relative}, a different configuration of the superpositions \cite{nguyen2019entanglement}, using higher dimensional quantum objects \cite{tilly2021qudits}, or using different measurement bases~\cite{yi2021massive}. Furthermore, Refs.~\cite{christodoulou2019possibility,Christodoulou_2020} propose that the QGEM experiment may be used to gain insights into the Planck mass and the discreteness of time, and possibly probe non-local gravitational interaction~\cite{marshman2020locality,Biswas:2005qr,Biswas:2011ar}. The non-local gravitational interaction tends to weaken the entanglement witness and the entanglement entropy~\cite{marshman2020locality}.

In this paper, we will explore a new design with 3 masses, each with an embedded qubit (instead of 2 masses with 2 qubits as proposed in the original QGEM protocol \cite{bose2017spin}). 
As we will witness the entanglement purely by measuring the qubits, we will refer to the previous and our current protocols henceforth as "2-qubit" and "3-qubit" protocols respectively.
We will show that this 3-qubit protocol 
%In this paper, we will explore a new design with 3 qubits (instead of 2 qubits as proposed in the original QGEM protocol~\cite{bose2017spin}). We will show that this setup 
performs better at generating entanglement, in particular when the decoherence effects are taken into account. However, this comes at the cost of requiring more measurements to characterise entanglement with a good enough level of certainty. In section \ref{sec:setup} different possible setups for a 3-qubit QGEM experiment will be discussed. The optimal setup will be identified by comparing the rate of entanglement generation in section \ref{sec:EE}.
In section \ref{sec:witness} we will discuss the witness expectation value for the different setups.
In section \ref{sec:decoherence} we will incorporate the decoherence in our analysis, to test how robust the different setups are to different decoherence rates.
Section \ref{sec:measurements} will show experimental simulations and discussions on the number of measurements needed to characterise the entanglement.
We will consider possible improvements by switching to qudits instead of qubits and discuss the merits of this approach in section \ref{sec:qudits}.

A short summary of our paper is that we find that the 3-qubit parallel setup (see figure \ref{fig:setup}) is better than other configurations with 3 qubits including the decoherence effects taken into account. Moreover, the entanglement entropy is dependent on the chosen subsystem.
Taking the partial trace of the \textit{middle} qubit gives the largest entanglement entropy. Similarly, when determining the witness, choosing the \textit{middle} qubit as the subsystem provides an improved witness. The 3-qubit setup outperforms the 2-qubit setup in generating entropy within $5 \, \si{\s}$ and it has a better witness. Furthermore, entanglement can be measured up to higher decoherence rates, and the number of measurements needed to confirm entanglement at higher decoherence rates $\gamma > 0.08 \, \si{\hertz}$ is similar to or smaller than for the 2-qubit setup.

%%%------------------------------------------------
\section{New 3-qubit QGEM setup} \label{sec:setup}

\begin{figure*}
\centering
\begin{subfigure}[ht]{0.25\linewidth}
\begin{tikzpicture}
\draw[black, dashed, thick] (0,-1) -- (0,-0.1);
\draw[black, dashed, thick] (1,-1) -- (1,-0.1);
\draw[black, dashed, thick] (2,-1) -- (2,-0.1);
\draw[black, thick] (0,-1.7) -- (0,-1.9) |- (0,-1.8) -- (1,-1.8) node[pos=0.5, below] {$d$} -| (1,-1.7) -- (1,-1.9);
\draw[black, thick] (-0.6,-1) -- (-0.4,-1) |- (-0.5,-1) -- (-0.5,0) node[pos=0.5, left] {$\Delta x$} -| (-0.6,0) -- (-0.4,0);
\filldraw[black]
(0,-1) circle (2pt) node[align=center, below] {\hspace{2mm}$\ket{0}_1$}
(1,-1) circle (2pt) node[align=center, below] {\hspace{2mm}$\ket{0}_2$}
(2,-1) circle (2pt) node[align=center, below] {\hspace{2mm}$\ket{0}_3$};
\draw[black]
(0,0) circle (2pt) node[align=center, above] {\hspace{2mm}$\ket{1}_1$}
(1,0) circle (2pt) node[align=center, above] {\hspace{2mm}$\ket{1}_2$}
(2,0) circle (2pt) node[align=center, above] {\hspace{2mm}$\ket{1}_3$};
\end{tikzpicture}
\caption{}
%\caption{The parallel setup of the three QGEM experiment where the spatial superpositions are aligned parallel to each other. Note that $d_\text{min} = d < \Delta x$.}
\label{fig:setup}
\end{subfigure}
\hfill
%\vspace{5mm}
%%%%%
\begin{subfigure}[ht]{0.4\linewidth}
\centering
\begin{tikzpicture}
\draw[black, dashed, thick] (-3,0) -- (-2.1,0);
\draw[black, dashed, thick] (-0.5,0) -- (0.4,0);
\draw[black, dashed, thick] (2,0) -- (2.9,0);
% \draw[black, thick] (-3,0.7) -- (-3,0.9) |- (-3,0.8) -- (-0.5,0.8) node[pos=0.5, above] {$d$} -| (-0.5,0.7) -- (-0.5,0.9);
% \draw[black, thick] (-3,0.2) -- (-3,0.4) |- (-3,0.3) -- (-2,0.3) node[pos=0.5, above] {$\Delta x$} -| (-2,0.2) -- (-2,0.4);
\draw[black, thick] (-3,-1.3) -- (-3,-1.5) |- (-3,-1.4) -- (-0.5,-1.4) node[pos=0.5, below] {$d$} -| (-0.5,-1.3) -- (-0.5,-1.5);
\draw[black, thick] (-3,-0.7) -- (-3,-0.9) |- (-3,-0.8) -- (-2,-0.8) node[pos=0.5, below] {$\Delta x$} -| (-2,-0.7) -- (-2,-0.9);
\filldraw[black] 
(-3,0) circle (2pt) node[align=center, below] {\hspace{2mm}$\ket{0}_1$}
(-0.5,0) circle (2pt) node[align=center, below] {\hspace{2mm}$\ket{0}_2$}
(2,0) circle (2pt) node[align=center, below] {\hspace{2mm}$\ket{0}_3$};
\draw[black]
(-2,0) circle (2pt) node[align=center, below] {\hspace{2mm}$\ket{1}_1$}
(0.5,0) circle (2pt) node[align=center, below] {\hspace{2mm}$\ket{1}_2$}
(3,0) circle(2pt) node[align=center, below] {\hspace{2mm}$\ket{1}_3$};
\end{tikzpicture}
\caption{}
%\caption{The linear setup of the three QGEM experiment where the spatial superpositions are aligned linearly to each other. Note that $d = d_\text{min} + \Delta x > \Delta x$.}
\label{fig:setup2}
\end{subfigure}
\hfill
%\vspace{5mm}
\begin{subfigure}[ht]{0.25\linewidth}
\centering
\begin{tikzpicture}
\draw[black, dashed, thick] (0.5,1) -- (-0.1,0.4);
\draw[black, dashed, thick] (1,1.7) -- (1,2.6);
\draw[black, dashed, thick] (1.5,1) -- (2.1,0.4);
\draw[black, thick] (0.5,0.6) -- (0.5,0.8) |- (0.5,0.7) -- (1.5,0.7) node[pos=0.5, below] {$d$} -| (1.5,0.6) -- (1.5,0.8);
\draw[black, thick] (0.6,1.7) -- (0.8,1.7) |- (0.7,1.7) -- (0.7,2.7) node[pos=0.5, left] {$\Delta x$} -| (0.6,2.7) -- (0.8,2.7);
\filldraw[black] 
(0.5,1) circle (2pt) node[align=center, right] {\hspace{0mm}$\ket{0}_1$}
(1,1.7) circle (2pt) node[align=center, right] {$\ket{0}_2$}
(1.5,1) circle (2pt) node[align=center, right] {\hspace{1mm}$\ket{0}_3$};
\draw[black]
(-0.2,0.3) circle (2pt) node[align=center, below] {\hspace{2mm}$\ket{1}_1$}
(1,2.7) circle (2pt) node[align=center, right] {$\ket{1}_2$}
(2.2,0.3) circle (2pt) node[align=center, below] {\hspace{2mm}$\ket{1}_3$};
\end{tikzpicture}
\caption{}
%\caption{The star setup of the three QGEM experiment where the spatial superpositions are aligned in a manner we label as 'star' pattern. Note that $d_\text{min} = d < \Delta x$.}
\label{fig:setup3}
\end{subfigure}
%\vspace{-5mm}
\caption[some caption]{The proposed setups of the QGEM experiment using $3$ qubits in spatial superpositions aligned in different configurations.
The superposition states of the $i^\text{th}$-qubit are denoted $\ket{0}_i$ and $\ket{1}_i$. 
The minimal distance $d_\text{min}=200 \, \si{\micro\m}$ between any two states and the superposition width $\Delta x = 250 \, \si{\micro\m}$ \cite{bose2017spin} is kept constant in all different setups.
The distance $d$ denotes the distance between any two neighbouring $\ket{0}$ states and is determined by the setup. (a) The parallel setup of the three QGEM experiment where the spatial superpositions are aligned parallel to each other. Note that $d_\text{min} = d < \Delta x$. (b) The linear setup of the three QGEM experiment where the spatial superpositions are aligned linearly to each other. Note that $d = d_\text{min} + \Delta x > \Delta x$. (c) The star setup of the three QGEM experiment where the spatial superpositions are aligned in a manner we label as 'star' pattern. Note that $d_\text{min} = d < \Delta x$.}
\label{fig:setup_all}
\end{figure*}
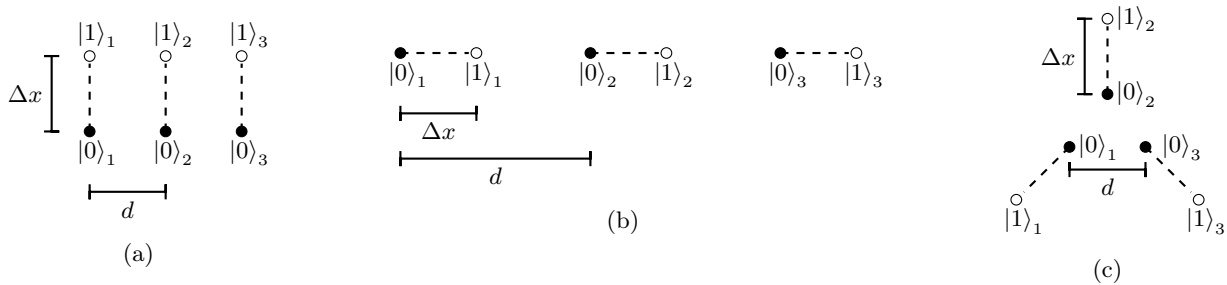

We first discuss the optimal setups for the 2- and 3-qubit cases\footnote{Throughout this paper the number of superpositions is denoted $n$ and the number of states in the superposition is denoted $D$. For spin systems, the value of $D=2s+1$, where $s$ is the spin state, for electronic spin $s=1/2$. Later we will consider qudits which have $D$ superposition states, but for now we consider only qubits ($D=2$).} %, which consist of two superposition states ($D=2$).}. 
Ref. \cite{nguyen2019entanglement} found that for the 2-qubit setup it is favourable to create the superpositions in the direction orthogonal to their separation (depicted in figure \ref{fig:setup}), as opposed to in the same direction as their separation which was proposed in the original paper \cite{bose2017spin} (depicted in figure \ref{fig:setup2}).
We will first introduce three different setups (figures \ref{fig:setup}-\ref{fig:setup3}), and in the next section we will study the generation of the entanglement in each setup to determine the better configuration for witnessing graviton induced entanglement.

For any 3-qubit setup the initial (unentangled) state is given as:~\footnote{Note that the different notations for the 3-qubit state used in this paper refer to equivalent states, that is $\bigotimes_{i=1}^3 \ket{0}_i = \ket{0}_1 \otimes \ket{0}_2 \otimes \ket{0}_3 = \ket{0_1 0_2 0_3}$. The state of the $i^\text{th}$ qubits is generally denoted by $j_i = 0,1$. For compactness and clarity different notation is used throughout this paper.}
\begin{equation}\label{eq:psi0}
    \ket{\Psi_0} 
    =
    \frac{1}{2\sqrt{2}} \bigotimes_{i=1}^{3} \left( \ket{0}_i + \ket{1}_i \right)
\end{equation}
The gravitational interaction between the two states is governed by the universal coupling $\sqrt{G}h_{\mu\nu}T^{\mu\nu}$, where the metric is $g_{\mu\nu}=\eta_{\mu\nu}+h_{\mu\nu}$ with $\mu,\nu=0,1,2,3$ and $h_{\mu\nu}$ is the perturbations around the Minkowski background $\eta_{\mu\nu}$. 
Due to the gravitational interaction each superposition states pick up a relative quantum mechanical phase $\phi \sim S/ \hbar \sim \frac{E\tau}{\hbar}$, where $S$ is the gravitational action.
This can be viewed as the result of the time evolution operator or equivalently, as a result of the Feynman path integral formalism for a particle moving through spacetime.
The interaction energy $E$ is determined by the gravitational potential energy $V$ which in the non-relativistic case is derived from the tree-level exchange of a virtual graviton \cite{marshman2020locality}.
The quantum origin of this potential in the non-relativistic limit of perturbative quantum gravity in the weak field regime is discussed in detail in Refs. \cite{marshman2020locality, Bose:2022uxe}.
The effect of higher order corrections will be negligible here.
%
% In Ref. \cite{Bose:2022uxe} the same entanglement phase is recovered from perturbative quantum gravity.
% It is shown that the presence of the massive systems gives rise to a change in the graviton vacuum energy and that from this change the interaction that entangles the masses emerges.
% In the static limit where the graviton is canonically quantised in a weak field regime the change in the gravitational vacuum energy is shown to be the same as Newton's potential energy at second order in the perturbation theory.
% The fact that this shift in gravitational vacuum energy is operator-valued also shows that it is a quantum rather then classical entity that governs the entangling interaction.

The total gravitational potential energy is given by $\hat{V}_\text{tot} = \hat{V}_{12} + \hat{V}_{23} + \hat{V}_{13}$, where the subscripts denote the interactions between the subsystems. 
e.g. $\hat{V}_{ij}$ is the potential energy between system $i$ at $x_i$ and system $j$ at $x_j$, which is given by $\frac{G m_i m_j}{\abs{\hat{x}_i - \hat{x}_j}}$.
%The potential in the non-relativistic limit is the same as that of Newtonian gravity, and seemingly classical. However, any possible gravitational candidate must reproduce the known behaviour at tested regimes. As discussed above, in QM and in quantum field theory we can recover the same potential in the second order perturbation theory by considering that the gravitational force between two states is being mediated by exchange of a mediator, i.e. virtual graviton state carrying 6 off-shell degrees of freedom.

We now give the system's state after the qubits have gravitationally interacted.
To avoid writing out all the terms, we will use the shorthand notation where  $j_i=0,1$ denotes the state of the $i^\text{th}$ qubit (so $\ket{j_i}$ is either $\ket{0}$ or $\ket{1}$).
\begin{equation}\label{eq:psit}
    \ket{\Psi(t=\tau)}
    =
    \frac{1}{2\sqrt{2}} \sum_{j_1,j_2,j_3=0,1} e^{i \phi_{j_1 j_2 j_3} \tau} \ket{j_1 j_2 j_3},
\end{equation}
with $\phi_{j_1 j_2 j_3}\tau$ the phase picked up due to the interaction via gravity between the systems during a time $\tau$ for the state $\ket{j_1 j_2 j_3}$.
Note that since $\phi_{j_1 j_2 j_3}$ is simply $U_\text{tot}/\hbar$, it is determined by the distance between the states $\abs{x_i - x_j}$.
Since the setup specifies the distance between two systems, the phases is specific to the setup.
For the parallel setup in figure \ref{fig:setup} we find:
\begin{equation}\label{eq:phase_def}
    \phi_{j_1 j_2 j_3}^{(\parallel)} = \frac{1}{\hbar} \sum_{i,k=1,2,3}^{i<k} \frac{G m^2}{\sqrt{d^2+(\Delta x (j_i - j_k))^2}} \, .
\end{equation}
The superscript $\parallel$ specifies the parallel setup. 
For the linear setup in figure \ref{fig:setup2} (denoted $-$) we find:
\begin{equation}\label{eq:phase_def2}
    \phi_{j_1 j_2 j_3}^{(-)} = \frac{1}{\hbar} \sum_{i,k=1,2,3}^{i<k} \frac{G m^2}{(k-i)d+\Delta x (j_k - j_i)} \, ,
\end{equation}
and for the star setup (denoted $*$) we find:
\begin{equation}\label{eq:phase_def3}
    \phi_{j_1 j_2 j_3}^{(*)} = \frac{1}{\hbar} \sum_{i,k=1,2,3}^{i<k} \frac{G m^2}{r_{ik}} \,,
\end{equation}
where $r_{ik}$ is the distance between the superposition instances labelled $i$ and $k$ in the star setup:
\begin{align}
    r_{ik} = &(1-j_i j_k) d + j_i j_k (d + \sqrt{\Delta x}) \nonumber \\
    &+ (j_k-j_i) \left(\sqrt{d^2 + \sqrt{3}d \Delta x + \Delta x^2}-d\right).
\end{align}
The expressions for the distances were found using trigonometry.
The distance $d$ between two neighbouring $\ket{0}$ states is determined by the superposition width $\Delta x = 250 \, \si{\micro\m}$ and by requiring a minimal distance between any two qubits $d_{\text{min}} = 200 \, \si{\micro\m}$.
This minimum distance is introduced  to ensure that the Casimir-Polder potential is negligible \cite{bose2017spin}.
Due to the chosen parameters, the separation between the two qubits in the parallel setup is $d=200 \, \si{\micro\m}$, and for the linear setup it is $d=450 \, \si{\micro\m}$, and for the star setup the edge of the inner triangle is $d = 200 \, \si{\micro\m}$ \footnote{The considered setups of figure \ref{fig:setup_all} are symmetrical in $d$ (and $m$). Considering asymmetric configurations where one of the distances is increased above $d_{\text{min}}$ is not favourable since it decreases the interaction which goes as $1/r$. (Similarly considering asymmetric configurations where one of the masses is smaller than what is considered the largest possible mass $m$ is not favourable since it decreases the interaction strength.)}.
Furthermore, we will consider the masses of $m \sim 10^{-14} \, \si{\kg}$ and an interaction time of $\tau_\text{int} = 2.5 \, \si{\s}$~\footnote{The interaction time of $\tau_\text{int} = 2.5 \, \si{\s}$ was chosen because it is both feasible experimentally and during this time the systems can become entangled enough to be detectable \cite{marshman2020locality,margalit2021realization}. If the interaction time were decreased, we will see in figures \ref{fig:entcomp}, \ref{fig:EW_time_0.1} and \ref{fig:EW_time_0.15} that the 3-qubit setup still generates more entanglement and provides a better witness.}.
These parameters fall within the feasible range discussed in \cite{bose2017spin,marshman2021large}.

From eq. \eqref{eq:psit} we find that the density matrix of the system is: $\rho(\tau) = \ket{\Psi(\tau)}\bra{\Psi(\tau)}$. 
\begin{equation}\label{eq:rhotau}
    \rho(\tau)
    =
    \frac{1}{8} \sum_{\substack{j_1,j_2,j_3=0,1\\j_1',j_2',j_3'=0,1}}
    e^{i (\phi_{j_1 j_2 j_3} - \phi_{j_1' j_2' j_3'})\tau} \bigotimes_{i,i'=1,2,3}  \ket{j_i} \bra{j_{i'}'}.
\end{equation}
For more on the density matrix formalism, see \cite{schlosshauer2007decoherence}.
With eq. \eqref{eq:rhotau}, we can analyse and compare the rate of entanglement generation for the different setups, which we will do by studying the entanglement entropy.

%%%------------------------------------------------
\section{Entanglement entropy test}\label{sec:EE}

As an initial test for comparing the 2-qubit setup with the 3-qubit setup, we assess the rate of entangement generation  for each version of the experiment. To do so, we will rely on the entanglement entropy (Von Neumann entropy) which measures the overall mixing of one of the subsystems of the 2 or 3-particle quantum state.

Using the density matrix formalism, the entanglement entropy can be found as \cite{von2018mathematical,nielsen2002quantum}: 
\begin{equation}
    S(\rho_a) = - \Tr(\rho_a \log_2(\rho_a)) = - \sum_i \lambda_i \log_2(\lambda_i) \, ,
\end{equation}
where $\rho_a$ is the partial density matrix of the subsystems $a$ with the eigenvalues $\lambda_i$.
The partial density matrix of a subsystem is found by taking the partial trace over the other subsystems, e.g. the partial density matrix describing the first system is denoted by $\rho_1 = \Tr_{2,3}(\rho)$, and can be used to find the entanglement entropy of the first subsystem, $S_1$. 
$\Tr_{2,3}$ denotes the partial trace over the subsystems $2$ and $3$. Although the entanglement entropy will be identical for either subsystem in the 2-qubit case ($S_1=S_2$), this is not the case when adding a third qubit as we now have several options for the system's partition. 

\begin{figure}[b]
    \centering
    \includegraphics[width=\linewidth]{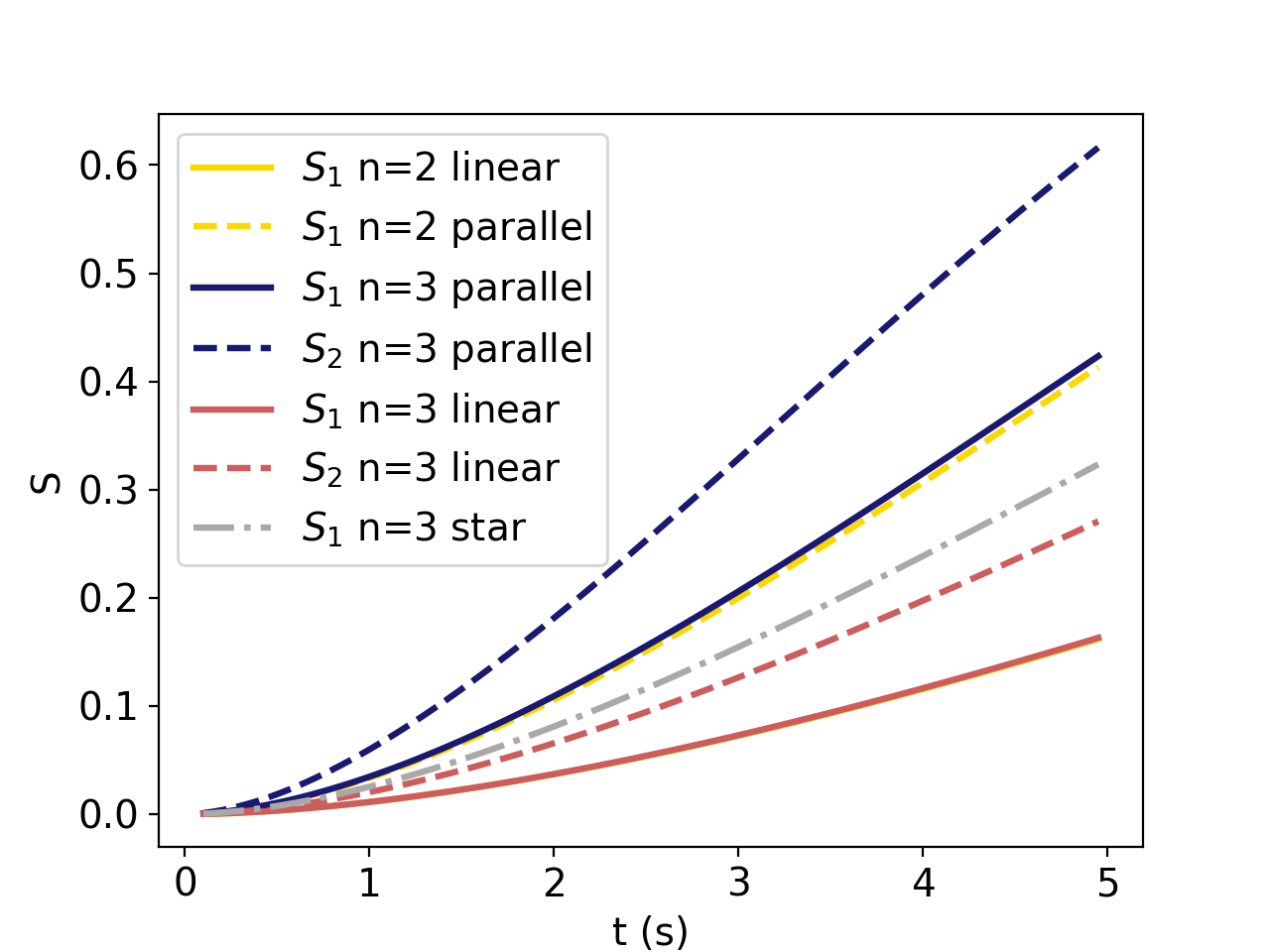}
    \caption{Comparison of the entanglement entropy generated within 5 seconds. Note that due to the symmetry of the setups, for the parallel and linear setup we have $S_1 = S_3$, and for the star setup we have $S_1 = S_2 = S_3$. During the first five seconds the $n=3$ parallel setup has the highest rate of entanglement generation compared to the other setups. The lines representing $S_1$ $n=3$ linear and $S_1$ $n=2$ linear almost overlap. The same plot for a larger time scale is given in appendix \ref{appendix:EE}.}
    \label{fig:entcomp}
\end{figure}

We compare the entanglement entropy for the 3-qubit with the 2-qubit system and find that an improvement is made for the parallel setup, especially for $S_2$, as can be seen from figure \ref{fig:entcomp}.
For the 2-qubit setup ($n=2$) we have considered the linear and parallel setups, these were discussed previously in \cite{bose2017spin,nguyen2019entanglement} respectively. 
Their state and density matrix can be derived from eq. \eqref{eq:rhotau} by leaving out the third particle $j_3$.
We have compared the 2-qubit ($n=2$) setups with the 3-qubit ($n=3$) parallel, linear and star setups (depicted in figures \ref{fig:setup}-\ref{fig:setup3} respectively).

From figure \ref{fig:entcomp}, it is clear that the setup that generates the most entropy is the parallel 3-qubit $S_2$. 
This setup considers the entropy between the middle subsystem (labelled 2) and the outer subsystems (labelled 1 and 3).
As noted above $S_2 > S_1 = S_3$, meaning the entanglement entropy is dependent on which systems are chosen to be traced out.
A likely explanation for these differences is that the average distance to the other subsystems is smaller for system 2 than for system 1 or 3 (see figure \ref{fig:setup}), and therefore it has a higher coupling to the other subsystems.
Note that due to the symmetries of the setup, for the linear and parallel case, we have  $S_1=S_3$, and for the star setup we have $S_1 = S_2 = S_3$.
From now on when considering the parallel or linear $n=3$ setup we will always consider the second subsystem $S_2$ (unless specified otherwise).

Of course, the use of entanglement entropy as a figure of merit for an experiment setup is not fully reliable. The introduction of noise in the experiment, and overall a realistic setting makes it impossible to distinguish mixed state from entanglement and the mixing resulting from the decoherence of the states. As such, following previous research on this experiment \cite{bose2017spin, nguyen2019entanglement, chevalier2020witnessing, tilly2021qudits}, we compare the different systems on the possibility of identifying the entanglement in a realistic setting by using the entanglement witness.

%%%------------------------------------------------
\section{Entanglement Witness} \label{sec:witness}

For the experimental detection of the entanglement one needs a witness, as discussed in~\cite{bose2017spin}.
The witness will be derived from a condition that determines when the states are separable or entangled. 
In Ref. \cite{bose2017spin}, the witness was derived from the witness for the Bell state. 
These type of witnesses are often not ideal in an experimental setup \cite{hyllus2005relations}.
%This resulted in the separability condition $(\phi_{01} + \phi_{10} - \phi_{00}-\phi_{11})\tau = 2\pi n$ with $n\in\mathbb{Z}$ and $\phi_{j_1 j_2}$ %as in eq. \eqref{eq:phase_def2} but with $i,k=1,2$ for the QGEM setup).
% \begin{equation}\label{eq:phase_def2}
%     \phi_{j_1 j_2} = \frac{\tau}{\hbar} \sum_{i,k=1,2}^{i<k} \frac{G m^2}{(k-i)d+\Delta x (j_k - j_i)},
% \end{equation}
%In \cite{chevalier2020witnessing}, it was shown that this witness fails when higher decoherence rates are introduced even in ideal settings.

Ref. \cite{chevalier2020witnessing} proposed a new witness based on the positive partial trace (PPT) criterion for separability of mixed states \cite{horodecki2001separability, peres1996separability}. In the 2-qubit case, this witness was found to work up to higher decoherence rates when decoherence was introduced into the model \cite{chevalier2020witnessing}.
In \cite{tilly2021qudits}, which considered the 2-particle QGEM  with qudits, the PPT witness was also found to be an optimal witness, and more efficient at detecting the entanglement than alternatives. For the 3-qubit setup explored in this paper, we will therefore also use the PPT witness.

The PPT witness gives the following criterion for the separability of the states: a state $\rho$ is separable iff its partial transpose is positive semi-definite, which is analogous to having no negative eigenvalues \cite{guhne2009entanglement}.
The partial transpose of the density matrix in eq. \eqref{eq:rhotau} is found to be:
\begin{equation}\label{eq:ppt}
    \rho^{T_2}(\tau)
    =
    \frac{1}{8} \sum_{\substack{j_1,j_2,j_3=0,1\\j_1',j_2',j_3'=0,1}}
    %\sum_{j_1,j_2,j_3, j_1',j_2',j_3'=0,1} 
    e^{i (\phi_{j_1 j_2' j_3} - \phi_{j_1' j_2 j_n'})\tau} \bigotimes_{i,i'=1,2,3}  \ket{j_i} \bra{j_{i'}'}
\end{equation}
where the superscript $T_2$ denotes the partial transpose of the second system is taken (similarly to $S_2$, transposing the middle subsystem provides a better witness because of the chosen partition).
Note that since the phase $\phi = \phi^{(\parallel)}, \phi^{(-)}, \phi^{(*)}$ differs for the different setups in figure \ref{fig:setup_all}, the witness will also be different for each of the three setups. According to the PPT criterion, the system is entangled if the PPT density matrix (eq. \eqref{eq:ppt}) has negative eigenvalues.

We will use this criterion to construct the entanglement witness $\mathcal{W}$, which is an observable that provides a sufficient (but not always necessary) condition to detect entanglement.
We construct the following witness \cite{horodecki2001separability}:
\begin{equation}
    \mathcal{W} = \ket{\lambda_-}\bra{\lambda_-}^T,
\end{equation}
where $\ket{\lambda_-}$ is the eigenvector corresponding to the smallest eigenvalue of $\rho^T$ (the partial transpose of $\rho$).
All separable states $\rho$ then satisfy \cite{horodecki2001separability}:
\begin{equation}
    \Tr{\mathcal{W}\rho} \geq 0\,.
\end{equation}
If $\Tr{\mathcal{W}\rho} < 0$, the state $\rho$ has negative eigenvalues and is therefore an entangled state.
Note that for the 3-qubit case, the witness is a sufficient but not a necessary condition for the entanglement generation, meaning that if $\Tr{\rho\mathcal{W}}\geq0$, the state can be either separable (not entangled) or non-separable (entangled).
An example of such a 3-qubit entangled state with $\Tr{\rho\mathcal{W}}\geq0$ is given in \cite{ha2016construction}.
The exact expressions for the witness operators for the 2- and 3-qubit setups are given in appendix \ref{appendix:EW} in terms of Pauli observables. 

\begin{figure}[t]
    \centering
    \includegraphics[width=\linewidth]{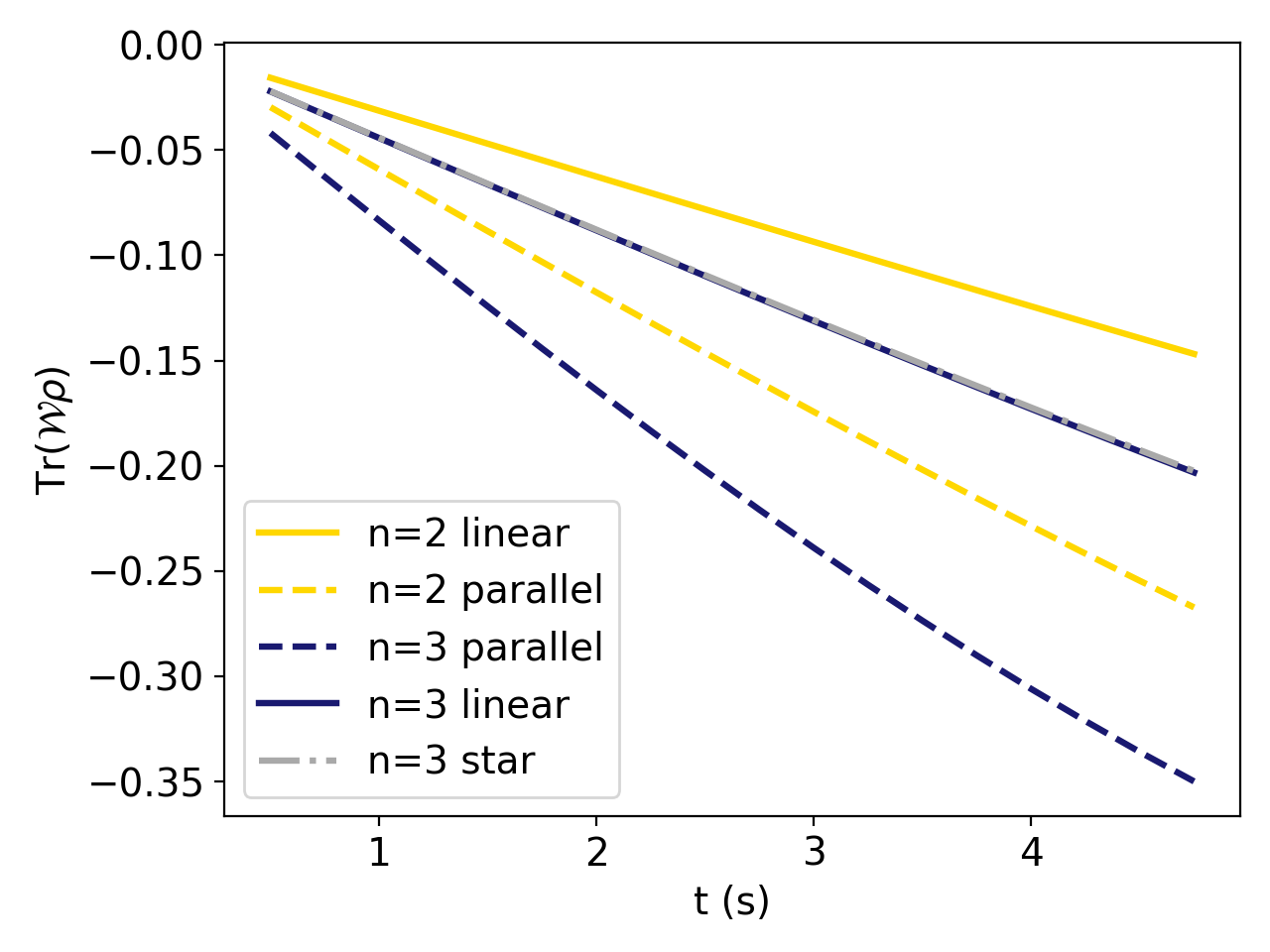} %from some_plot.py
    \caption{The expectation value of the PPT witness $\mathcal{W}$ with respect to time. For the 3-qubit cases we have considered $\rho^{T_2}$, meaning that system 2 is transposed as in eq. \eqref{eq:ppt}. The lines representing the $n=3$ linear and the star setups approximately overlap.
    We see that in the chosen basis the $n=3$ parallel setup provides the best (most negative) witness.}
    \label{fig:witness_over_time}
\end{figure}

The expectation values of the PPT entanglement witness for the setups discussed in section \ref{sec:setup} are shown in figure \ref{fig:witness_over_time}.
The 3-qubit parallel setup does very well compared to the other setups, which makes us optimistic about its potential for detecting the entanglement.
Since the witness is basis-dependent, one can only compare the shapes, and should be careful in comparing the values. 
Nevertheless, figure \ref{fig:witness_over_time} shows a much faster decrease in the entanglement witness expectation value for the parallel 3-qubit setup compared to the other setups.

Although adding a qubit seems to yield a better witness it also comes with a more difficult experimental setup. 
We can decompose the witness in terms of Pauli matrices, which can be measured experimentally. 
Examples of these decompositions for two and three qubits are shown in the appendix \ref{appendix:EW}.

\begin{table}[h!]
\centering
\begin{center}
\setlength{\tabcolsep}{10pt} % Default value: 6pt
\renewcommand{\arraystretch}{1.5}
\begin{tabular}{||c c c c||} 
 \hline
 n & setup & \#operators & \#operator groups\\ [0.5ex] 
 \hline\hline
 2 & par & 4 & 3 \\ 
 \hline
 2 & lin & 9 & 8 \\
 \hline
 3 & par & 26 & 12 \\
 \hline
 3 & lin & 47 & 22 \\ 
  \hline
 3 & star & 56 & 26 \\ [1ex]
 \hline
\end{tabular}
\end{center}
\caption{Number of Pauli operators that make up the decomposed PPT witness and the number of Pauli operator groups for the different setups. Operator groups are formed by operators which can be measured together and can be found using the Largest Degree First Colouring (LDFC) algorithm. The partial transpose is taken such that the witness is optimal. For the 3-qubit case the second system is partial transposed as in eq. \eqref{eq:ppt}}
\label{table:op_grouping2}
\end{table}

In general, given each particle can be measured in one of three Pauli basis elements (plus identity), there is a total  $\mathcal{O}(4^n)$ terms to be measured for an entanglement witness acting on $n$ particles. The number of Pauli matrices that need to be measured scales exponentially with the number of qubits, therefore, increasing exponentially the number of measurements required to characterise the entanglement.

%%%
Given that we aim to produce the experimental setup that is most efficient in detecting the entanglement, this could become a large impediment to using three or more qubits. As an illustration, while the 2-qubit setup only requires the measurement of three operators to construct the witness, the 3-qubit setup requires 25 operators to be measured. 

To address this issue, we will turn towards the literature regarding grouping a list of Pauli matrices into Abelian (commutative) groups, traditionally used in quantum computing (and initially stemming from error correction theory). Pauli matrices that commute with each other can be jointly diagonalised by basis rotation \cite{griffiths2005introduction}, allowing them to be jointly measured. Grouping of terms for the QGEM experiment with qudits was proposed in \cite{tilly2021qudits} by identifying groups based on general commutativity of the operators \cite{Yen2020, Hamamura2020, Gokhale2019_short}. 
However, authors noted that the realisation of the joint measurements by following this type of grouping may require non-local operations~\footnote{By non-local operations, we mean operations that act on several qubits and which cannot be applied to each qubit separately - these include for instance entangling operations or swap operations \cite{Gokhale2019_long, Crawford2021} which might be difficult to implement in practice.}. 
These techniques are widely used in the context of early-stage quantum computation to reduce the number of measurements required to perform methods such as the Variational Quantum Eigensolver, see \cite{Peruzzo2014}. 
%%%
Here we propose to group Pauli matrices based on Qubit-Wise commutation \cite{McClean2016, Kandala2017, Hempel2018, Rubin2018, Kokail2019, Gokhale2019_long} rather than the general commutation (i.e. we group two Pauli matrices together if each qubit-operator commute in the first matrix commutes with the respective qubit-operator in the second matrix) \cite{Gokhale2019_short, Gokhale2019_long, Hamamura2020, Yen2020}. 
While this in general means that fewer savings can be achieved in terms of the total number of measurements we can perform, these joint measurements can be realised through local operations only. It is worth noting as well that such grouping strategies could result in additional sampling noise from the appearance of co-variance between the operators being measured jointly \cite{McClean2016}. These were shown however to be the exception rather than the rule \cite{Gokhale2019_long}, and therefore we will leave the detailed analysis of these possible co-variances to future analyses. 
%%%
To perform the grouping, we use the Largest Degree First Colouring (LDFC) algorithm \cite{Welsh1967}, though for a small number of particles this can be done manually. LDFC is a traditional heuristic algorithm for graph colouring. It has been shown at least in some examples to perform somewhat better than other colouring heuristics in the context of Pauli string grouping \cite{Crawford2021} - a description of the method in this context can be found in the supplementary materials of \cite{Hamamura2020}.
We can reduce the number of operators in the 3-particle case from $25$ to $12$ using this method. The groups we have found are presented as an example in appendix \ref{appendix:EW}. 
A comparison of the number of operators in the Pauli decomposition can be seen in table \ref{table:op_grouping2}. 
By grouping some of the operators together with the number of operators needed for the measurement of the witness significantly decreases. 
Due to the grouping of operators the cost for conducting the 3-qubit experiment in the parallel setup seems relatively cheap, it reduces the number of operators that need to be measured to witness the entanglement from $25$ to $12$.
The linear and star setups will not be considered in the remainder of this paper since they do not provide a better entanglement generation (see figure \ref{fig:entcomp}) nor a better witness (see figure \ref{fig:witness_over_time}), and require more operator groups/operators to construct the witness (see table \ref{table:op_grouping2}).
The number of measurements needed to characterise entanglement by measuring the witness increases as the decoherence is introduced into the setup.

%%%------------------------------------------------
\section{Decoherence} \label{sec:decoherence}

We have assumed so far that the system is completely unaffected by its environment.
Given our aim is to estimate which setup is more appropriate, it is necessary to simulate a realistic case where we must consider the effects of decoherence.
The interaction of the system with its environment causes the system to share, and subsequently potentially loose, this information information. This is defined as decoherence (see \cite{schlosshauer2007decoherence} for an introduction to decoherence theory).
Although the decoherence is determined by the specific interactions with the environment, we can still consider a general approach. 
Assuming that the environmental state $\ket{E_i}$ couples to the system's position states $\ket{\vec{x}(i)}$ \cite{schlosshauer2014quantum}, we rewrite eq. \eqref{eq:psi0} to describe both the environment and the system:
\begin{equation}\label{eq:rho_env}
    \ket{\Psi_0}
    = \frac{1}{2^{3/2
    }} \sum_{j_1,j_2,j_3=0,1} \ket{\vec{x}(j_1) \vec{x}(j_2) \vec{x}(j_3)} \ket{E_{j_1} E_{j_2} E_{j_3}} \, .
\end{equation}
We  have assumed that the qubits are independent of each other at $t=0 \, \si{\s}$, and that their coupling to the environment is independent \cite{van2020quantum}. The density matrix describing the environment and the system is found from eq. \eqref{eq:rho_env} as usual: $\rho(0) = \ket{\Psi_0}\bra{\Psi_0}$. 
We can extract the system's (s) entanglement by tracing out the environmental (e) degrees of freedom $\rho_s = \Tr_e(\rho) = \sum_i \bra{E_i} \rho \ket{E_i}$. We find:
\begin{align}\label{eq:rhodeco2}
    &\rho_S(0)
    =
    \frac{1}{8} \sum_{\substack{j_1,j_2,j_3=0,1}} \ket{j_1 j_2 j_3} \bra{j_1 j_2 j_3} \nonumber \\
    &+
    \frac{1}{8} \sum_{\substack{j_1,j_2,j_3=0,1\\j_1',j_2',j_3'=0,1}}^{j_1 j_2 j_3 \neq j_1' j_2' j_3'} \ket{j_1 j_2 j_3} \bra{j_1' j_2' j_3'}\bra{E_{j_1} E_{j_2} E_{j_3}}\ket{E_{j_1'} E_{j_2'} E_{j_3'}}
\end{align}
The above eq. \eqref{eq:rhodeco2} shows that the system loses coherence as $\bra{E_{j_1 j_2 j_3}}\ket{E_{j_1' j_2' j_3'}} \to 0$. % \footnote{As the system loses coherence, the off-diagonal terms in the density matrix (also called the coherence terms) go to zero. In eq. \eqref{eq:rhodeco2} the off-diagonal terms are given in the second line.}. 
As is done in many decoherence models \cite{schlosshauer2014quantum}, we assume that the overlap between the environment states decreases exponentially over time with a rate $\gamma$, known as the decoherence rate.
\begin{equation*}
    \bra{E_{j}(t)}\ket{E_{j'}(t)} \propto e^{-\gamma t} \qq{} j\neq j'
\end{equation*}
Clearly at $t=0$ the states have not lost coherence.
By time-evolving eq. \eqref{eq:rhodeco2}, we obtain:
\begin{align}\label{eq:rhodeco}
    &\rho_S(\tau)
    =
    \frac{1}{8} \sum_{\substack{j_1,j_2,j_3=0,1}} \ket{j_1 j_2 j_3} \bra{j_1 j_2 j_3}  \\
    &+
    \frac{1}{8} \sum_{\substack{j_1,j_2,j_3=0,1\\j_1',j_2',j_3'=0,1}}^{j_1 j_2 j_3 \neq j_1' j_2' j_3'} e^{-\delta \gamma \tau}
    e^{i(\phi_{j_1 j_2 j_3} - \phi_{j_1' j_2' j_3'})\tau} \ket{j_1 j_2 j_3} \bra{j_1' j_2' j_3'} \nonumber
\end{align}
with $\phi_{j_1j_2j_3}$ as given in eq. \eqref{eq:phase_def}, and $\delta = 3 - \delta_{j_1 j_1'} - \delta_{j_2 j_2'} - \delta_{j_3 j_3'} \in [1,2,3]$.
The exponential decay term causes the off-diagonal terms to go to zero over time, making it impossible to measure the entanglement, and progressively transforming the experimental quantum state into a maximally mixed state.
The presence of the environment provides a constraint on the experiment time $\tau$.

\begin{figure}[t]
    \centering
    \includegraphics[width=\linewidth]{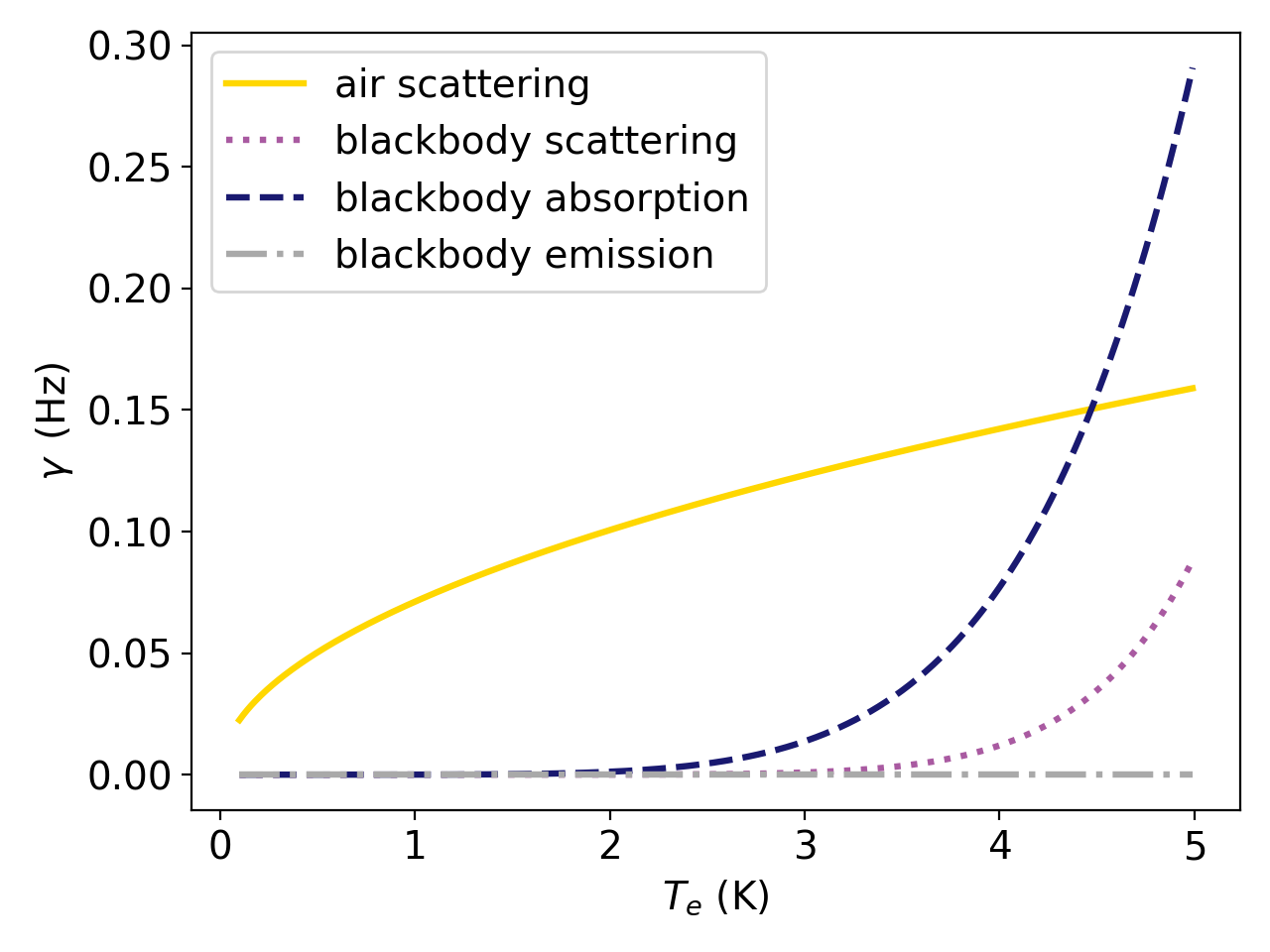}
    \caption{An estimate of the decoherence rate due to interaction with air molecules and blackbody photons. For environmental temperatures $T_e = 0.1 - 5 \, \si{\kelvin}$ the wavelength of the air molecules is much smaller ($1-15 \, \si{\nano\m}$) and the wavelength of the blackbody photons is much larger ($3-98 \, \si{\mm}$) than the superposition width ($250 \, \si{\micro \m}$).
    Therefore, we can use both the short and long wavelength limits to study their decoherence effects for this range of temperatures.
    The number density for the gas is taken to be $N/V = 10^8 \, \si{\m^{-3}}$,
    the radius of the superposition particles is $a = 10^{-6} \, \si{\m}$, and we have used the dielectric constant of a material similar to that of a diamond. 
    From the figure, we expect the decoherence rate of at least $0.5 \, \si{\hertz}$ for the ambient temperatures $T_e>0.5 \, \si{\kelvin}$. 
    The decoherence due to the blackbody emission is dependent on the internal temperature ($T_i = 0.15 \, \si{\kelvin}$), and therefore has a constant value of $\sim 10^{-10} \, \si{\hertz}$. See appendix \ref{appendix:decoherence_rate} for more details.
    }
    \label{fig:gamma_estimate}
\end{figure}

In appendix \ref{appendix:decoherence_rate}, the interaction with the environment has been studied in a more detailed fashion based on the analysis performed in \cite{van2020quantum}. The results of the calculations performed in this appendix are represented in figure \ref{fig:gamma_estimate}, which shows the estimated decoherence rate as a function of the ambient temperature.
We estimate that $\gamma>0.05 \, \si{\hertz}$ for an environmental temperature of at least $0.5 \, \si{\kelvin}$.

\begin{figure}[h!]
    \centering
    \includegraphics[width=\linewidth]{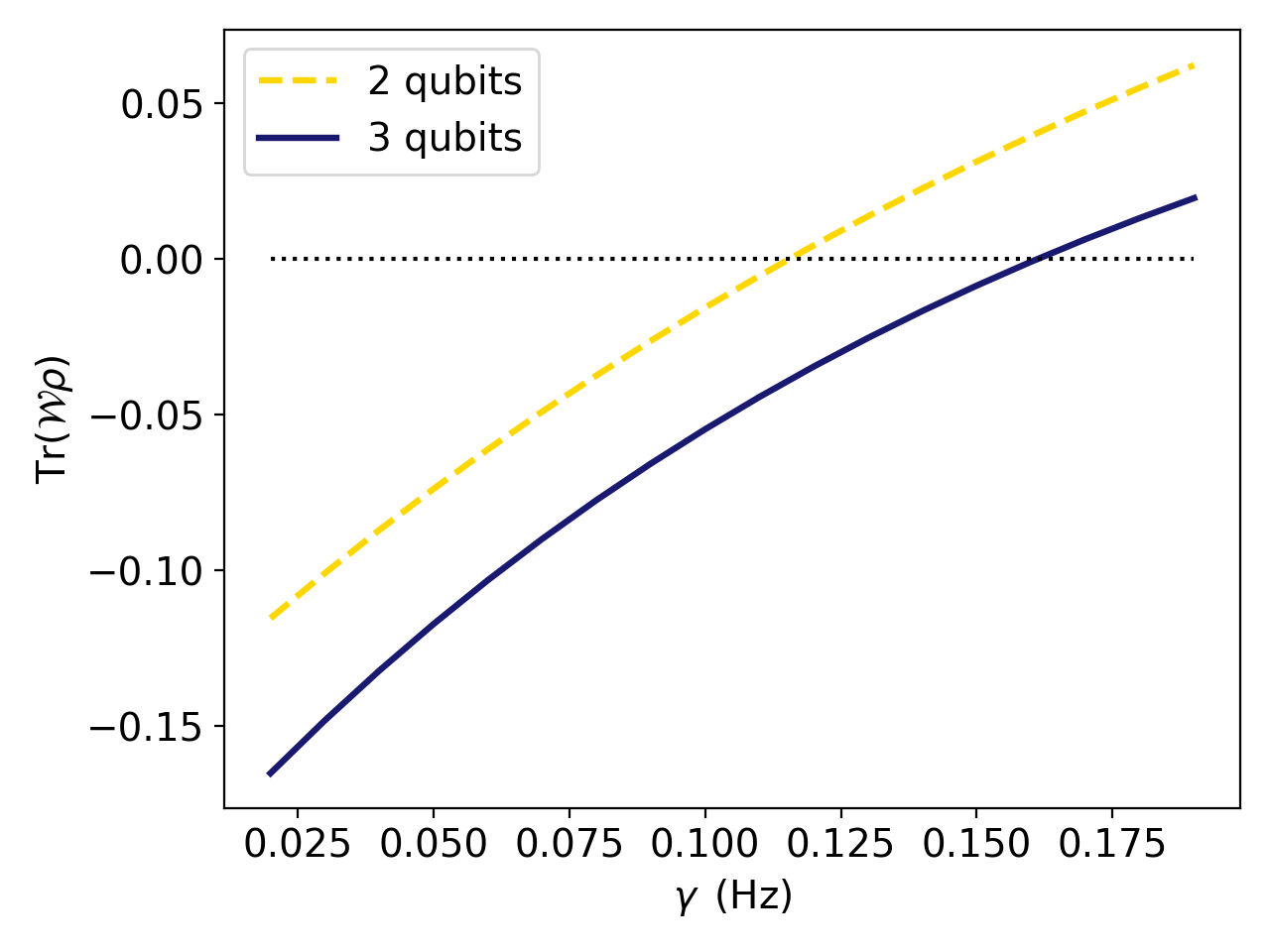} 
    % from new_qgem_model_EW_deco_add.py
    \caption{The PPT entanglement witness expectation value is plotted for the 2- and 3-qubit parallel setups, for $\gamma \in [0.02,0.20] \, \si{\hertz}$ and at $\tau=2.5 \, \si{\s}$. The partial transpose for the 3-qubit setup is taken over the second subsystem.
    The 3-qubit setup is negative for higher decoherence rates compared to the 2-qubit setup.
    In the appendix \ref{appendix:EE} we show that taking the partial transpose over the first subsystem does not give such an improvement in the witness. The dotted line indicates $\Tr(\mathcal{W}\rho) = 0$. }
    \label{fig:decoherencerates}
\end{figure}

We  have compared the witness expectation value for different decoherence rates $\gamma$ in figure \ref{fig:decoherencerates}.
However, due to the witness being basis-dependent this comparison should be considered with caution.
The 3-qubit parallel setup (where we take $\rho^{T_2}$) seems to be robust against decoherence compared to the 2-qubit setup. 
The PPT entanglement witness becomes positive at $\tau=2.5 \, \si{\s}$ around $\gamma>0.12 \, \si{\hertz}$ for $n=2$ and around $\gamma>0.16 \, \si{\hertz}$ for $n=3$. %However, due to the witness being basis-dependent this comparison should be considered with caution.
In figure \ref{fig:extra_EW} in appendix \ref{appendix:EE}, we have shown that finding of the witness by partially transposing the first qubit ($\rho^{T_1}$) does not improve the witness compared to the 2-qubit setup.

\begin{figure}[h!]
    \centering
    \includegraphics[width=\linewidth]{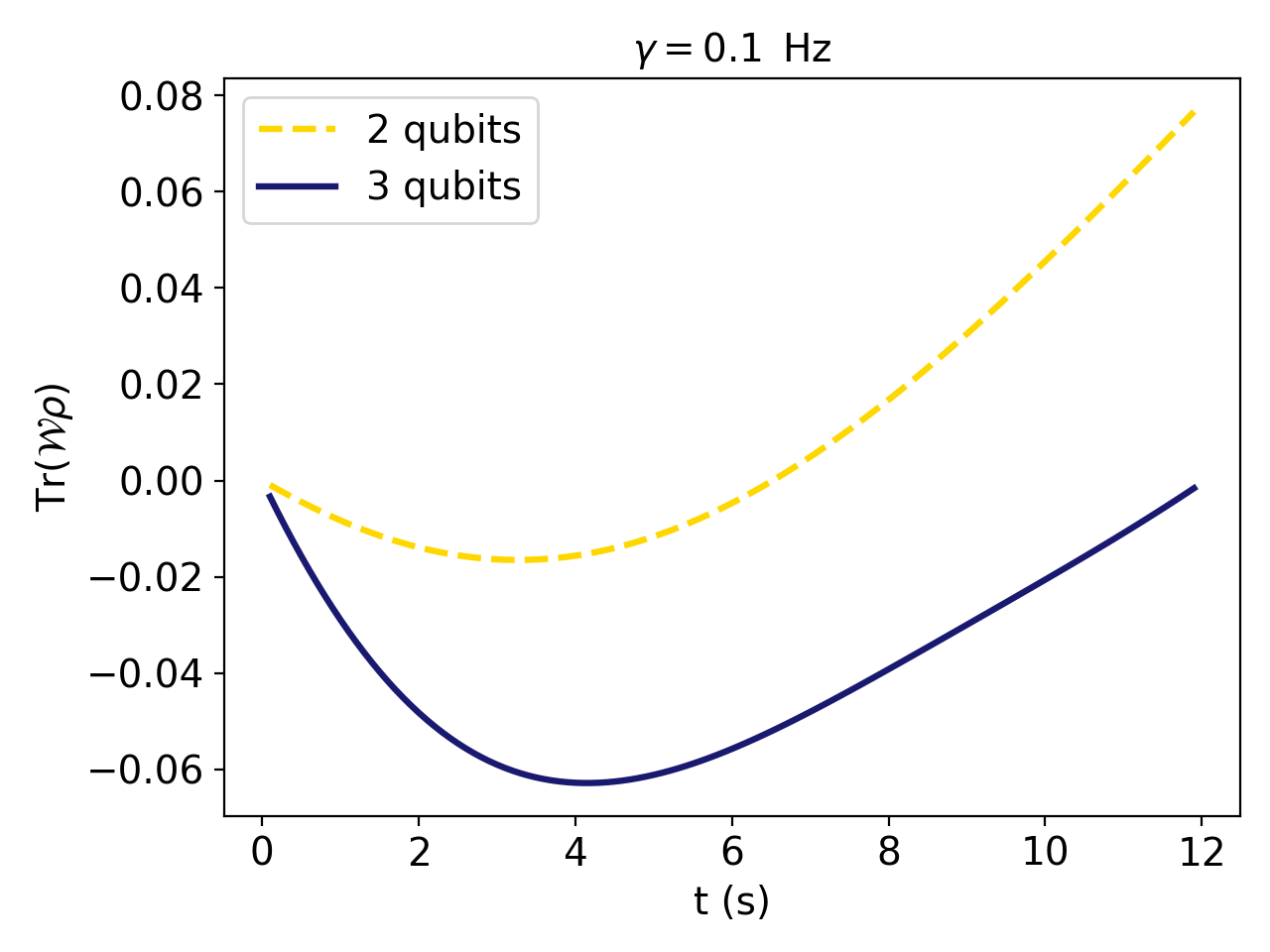}
    % from new_qgem_model_EW_deco_add.py
    \caption{The PPT entanglement witness expectation value over time for $\gamma=0.1 \, \si{\hertz}$ for 2 and 3 qubits in the parallel setup. The partial transpose for the 3-qubit setup is taken over the second subsystem. Due to the competition between the entanglement and the decoherence the graphs will become positive at some point when the decoherence effects become too strong. This will happen sooner for the 2-qubit setup than for the 3-qubit setup.}
    \label{fig:EW_time_0.1}
\end{figure}

\begin{figure}[t]
    \centering
    \includegraphics[width=\linewidth]{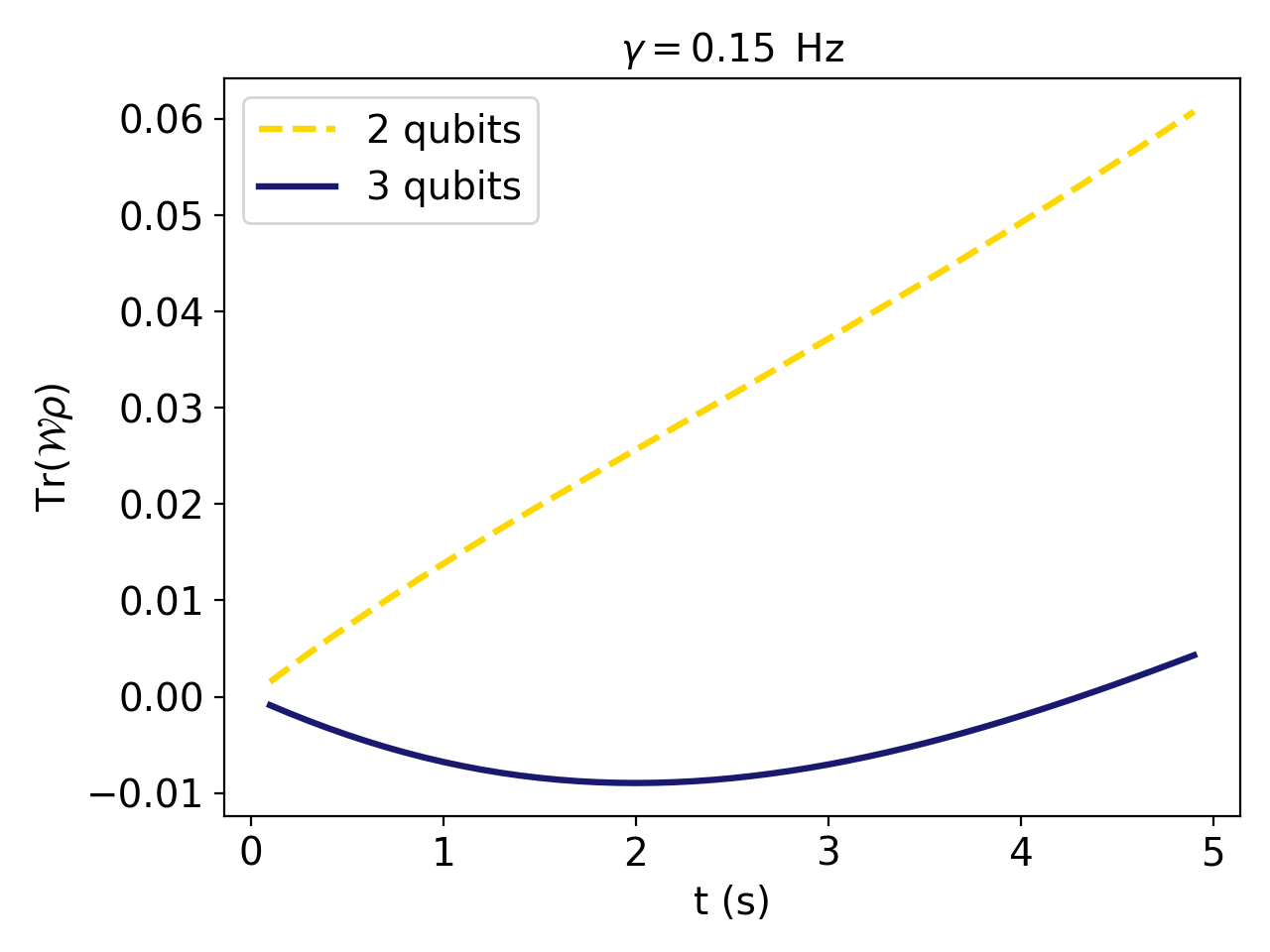}
    \caption{The PPT entanglement witness expectation value over time for $\gamma=0.15 \, \si{\hertz}$ for 2 and 3 qubits in the parallel setup. The partial transpose for the 3-qubit setup is taken over the second subsystem.
    The witness for the 2-qubit setup is never negative due to the strong decoherence effects.
    It is negative for about $4 \, \si{\s}$ for the 3-qubit setup. }%The dotted line indicates $\Tr(\mathcal{W}\rho) = 0$.}
    \label{fig:EW_time_0.15}
\end{figure}

Figures \ref{fig:EW_time_0.1} and \ref{fig:EW_time_0.15} show the entanglement witness as a function of time for $\gamma = 0.1 \, \si{\hertz}$ and $\gamma=0.15 \, \si{\hertz}$ respectively.
We see that the experiment time $\tau$ is important in protecting the system against decoherence.
Comparing the shapes of the lines in these figures also indicate that the 3-qubit setup will likely be more robust against decoherence since it has a steeper decrease.
From figure \ref{fig:EW_time_0.1}, we can see that if the experiment takes a sufficiently long time, the entanglement would not be measurable for any of the systems considered as the witness will become positive due to the decoherence effects.
In figure \ref{fig:EW_time_0.15}, the  entanglement is seen to be undetectable for the 2-qubit setup, and only detectable for the 3-qubit setup if $\tau < 4 \, \si{\s}$. The shapes of the curves in these two graphs is due to the competition between the entanglement and the decoherence.


Intuitively the 3-qubit setup is understood to be more robust against decoherence effects because of this competition between the entanglement of the subsystems and the decoherence.
%whereby a greater number of entangled states must be decohered before the states become separable.
Since the 3-qubit setup generates more entanglement between the subsystems within 5 seconds than the 2-qubit setup (see figure \ref{fig:entcomp}), it takes more decoherence (meaning either a longer time or a higher decoherence rate) to destroy this entanglement between the subsystems.
However, the fact that our 3-qubit system is more robust against decoherence is not a general argument since the entanglement generation depends not only on the number of particles but also on the setup and the choice of subsystem, as discussed in section \ref{sec:setup}.

Our 3-qubit setup looks very promising, but before drawing any conclusions we must simulate the number of measurements needed to witness the entanglement in a lab.

%%%------------------------------------------------
\section{Simulating QGEM experiment} \label{sec:measurements}

The experimental simulations (with which we mean a numerical simulation of the experiment statistics) are done in the same way as in Ref.~\cite{tilly2021qudits}.
The expectation value and the standard error are computed by repeatedly measuring the quantum state resulting from the experiment against the Pauli operators.
For each fixed number of repeated measurements the confidence level of confirming that the state is entangled is found. 
A confidence interval ($CI$) for the expectation value of a witness $\mathcal{W}$ can be computed as follows \cite{tilly2021qudits, Dekking2005}: 
 \begin{equation}\label{eq:confidence_level}
    CI_\mathcal{W} = [\langle \mathcal{W}\rangle - t_{\alpha} s_{\mathcal{W}}, \langle \mathcal{W}\rangle  + t_{\alpha} s_{\mathcal{W}} ] \, ,
\end{equation}
with $t_{\alpha}$ the $t$-value corresponding to the desired level of confidence, and $s_{\mathcal{W}}$  is the standard error of the measurement population~\footnote{The t-value is a measure of the difference between the data ($ \langle\mathcal{W} \rangle$ in our case) and the null hypothesis ($\mu_0 = 0$ in our case). In general a larger $t$-value means that there is a more evidence against the null hypothesis. For details about statistical testing and analysis we recommend \cite{Dekking2005}.}. 
To estimate a given confidence level tested against the null hypothesis $\langle \mathcal{W} \rangle \ge \mu_0$, with $\mu_0 = 0$, we compute the $t$-values by performing a one-sided $t$-test~\footnote{A one-sided $t$-test, as opposed to a two-sided test, only performs statistical tests in one direction (instead of two). Since we are interested in rejecting the null hypothesis $\langle \mathcal{W} \rangle \ge 0$, a one-sided test is applicable.} \cite{tilly2021qudits, Dekking2005}: 
 \begin{equation}
    t =\frac{\lvert \langle \mathcal{W}\rangle  - \mu_0 \rvert }{s_{\mathcal{W}}}\,.
\end{equation}
The $t$-value provides us information about the probability of the expectation value $\mathcal{W}$ being below the $\mu_0$ (i.e. that we can reject the null hypothesis). To translate this information into a confidence level, we can look at a $t$-distribution table and recover the so-called $p$-value which corresponds to the probability of making an error when rejecting the null hypothesis. Therefore, $p$ is the probability that the true value of $\mathcal{W}$ is above $\mu_0 = 0$ despite  the data gathered. To get the confidence level, we just need to compute $1 - p$. For further details, please refer to \cite{Dekking2005}, or for another application to quantum observables, please refer to the appendix of \cite{tilly2021qudits}.
As discussed in section \ref{sec:witness}, if the system is not confirmed to be entangled, it only means that the entanglement was not measurable. 
It does not necessarily mean that the system is separable. 

\begin{figure}[t]
    \centering
    \includegraphics[width=\linewidth]{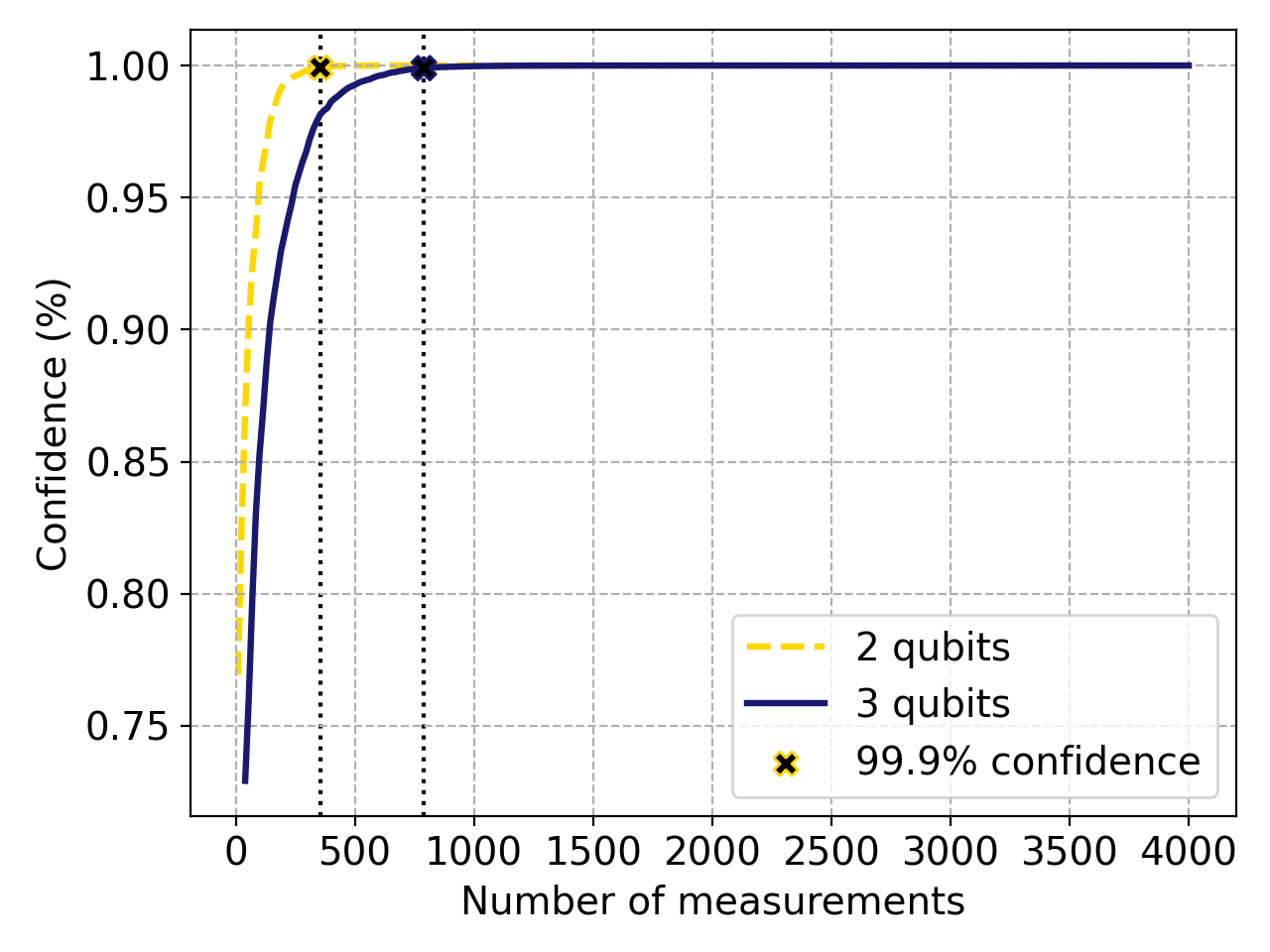}
    % from plot_test_data_n23.py
    \caption{Experimental simulation of the QGEM experiment for the 2-qubit and 3-qubit parallel setups at $\tau=2.5 \,\si{\s}$. The crosses indicate $99.9 \%$ confidence that the state is entangled.
    The witness expectation values are $-0.146$ for the 2-qubit setup, and $ -0.202$ for the 3-qubit setup (these can also be read off from the figure \ref{fig:decoherencerates}).
    }
    \label{fig:data_n26_15}
\end{figure}

\begin{figure}[h!]
    \centering
    \includegraphics[width=\linewidth]{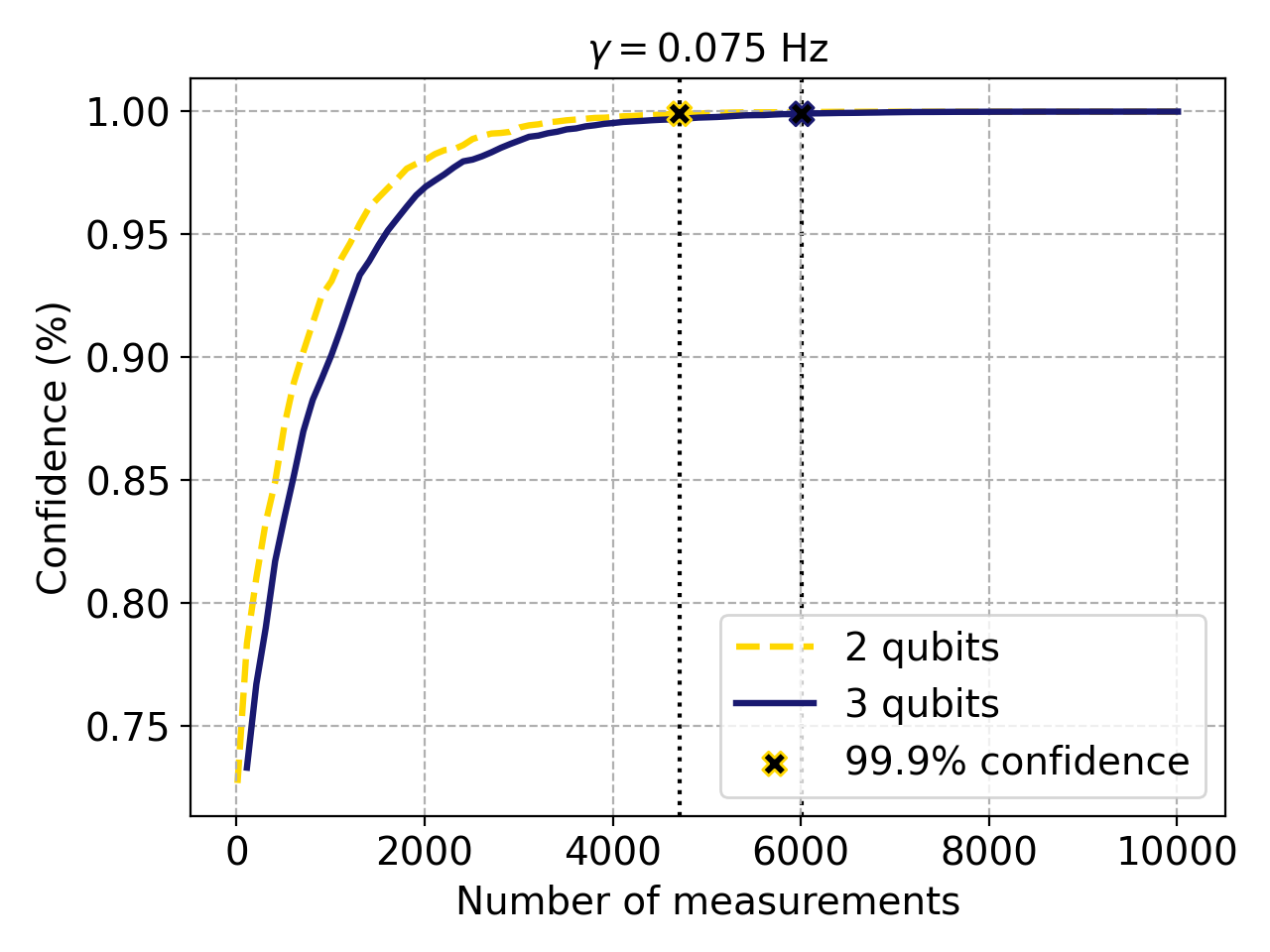} 
    %from Plot1_n2-6_gamma_0.075
    \caption{Experimental simulation of the QGEM experiment for the 2-qubit and 3-qubit parallel setups with $\gamma = 0.075 \, \si{\hertz}$ and at $\tau=2.5 \,\si{\s}$. 
    The crosses indicate  $99.9 \%$ confidence for the entanglement, this number increases due to the introduction of the decoherence into the model.
    The witness expectation values are $-0.043$ for the 2-qubit setups and $-0.084$ for the 3-qubit setup (see figure \ref{fig:decoherencerates}).}
    \label{fig:data_n26_deco_10}
\end{figure}

Figure \ref{fig:data_n26_15} compares the confidence levels given in eq. \eqref{eq:confidence_level} as a function of the number of measurements for the 2- and 3-qubit parallel setups after $\tau=2.5 \, \si{\s}$.
Without the decoherence, due to the extra qubit in the 3-qubit setup more measurements are needed to confirm with confidence ($99.9\%$) that the system is entangled. However, from the discussions in the previous section we expect that if the decoherence is included, at some point while increasing the decoherence rate, the 3-qubit setup will require a fewer number of measurements due to it being more resistant to the effects of decoherence.
Figure \ref{fig:data_n26_deco_10} compares the confidence levels given in eq. \eqref{eq:confidence_level} with a decoherence rate of $\gamma=0.075 \, \si{\hertz}$.  Due to the introduction of the decoherence, the number of measurements needed to confirm the entanglement increases.
\begin{figure}[t]
    \centering
    \includegraphics[width=\linewidth]{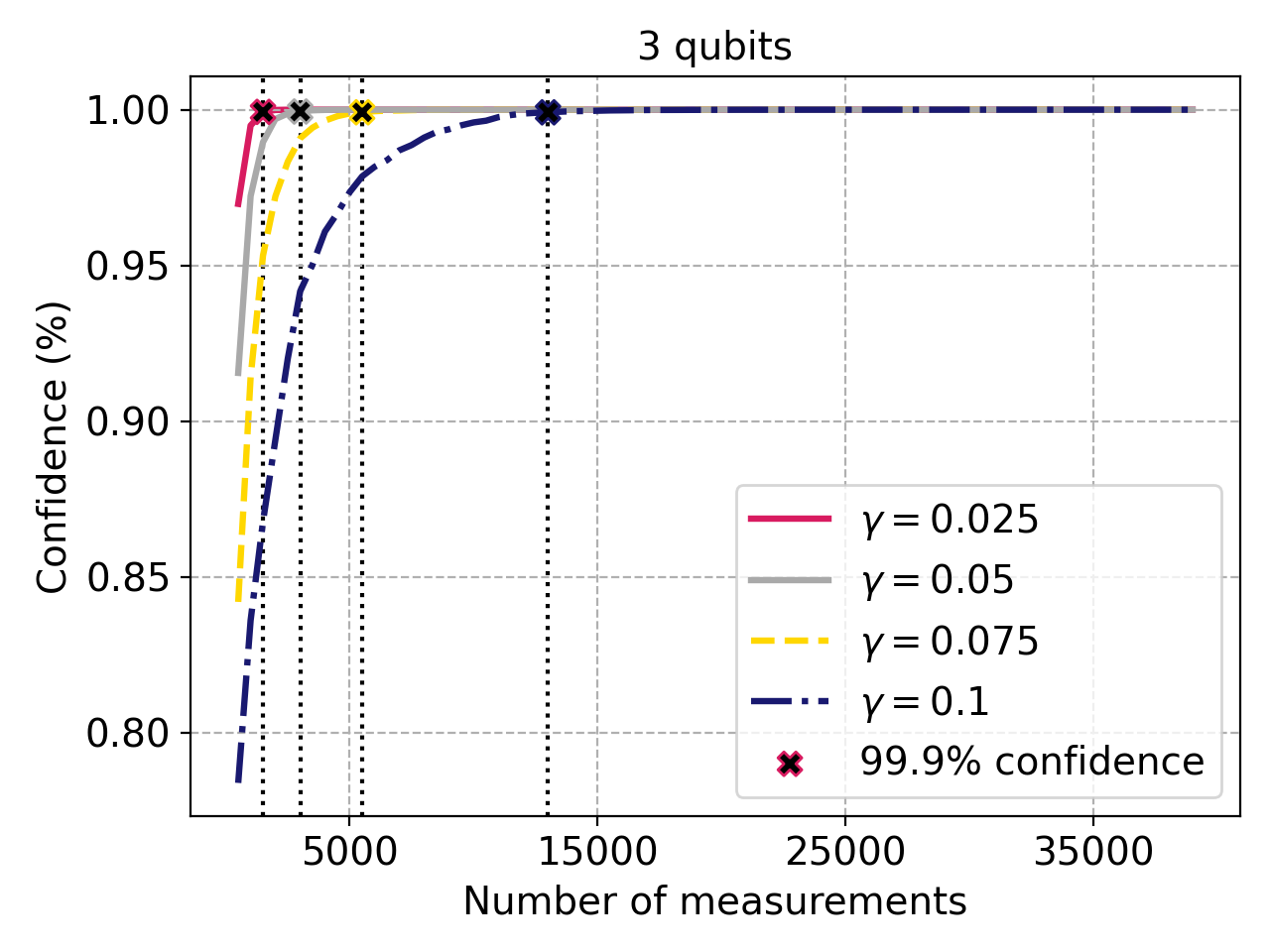}
    % from plot_test_data_ns_gammas.py
    \caption{Experimental simulation of the QGEM experiment for the 3-qubit parallel setup for different decoherence rates at $\tau=2.5 \,\si{\s}$. With $\gamma = [0.025, 0.050, 0.075, 0.1] \, \si{\hertz}$, respectively, having the witness expectation values $-0.156$, $-0.117$, $-0.084$, $-0.055$ (see figure \ref{fig:decoherencerates}). The number of measurements needed to confirm the entanglement with the confidence increases for higher decoherence rates, but not as much as for the 2-qubit setup.}
    \label{fig:data_n3_gamma}
\end{figure}
\begin{figure}[h!]
    \centering
    \includegraphics[width=\linewidth]{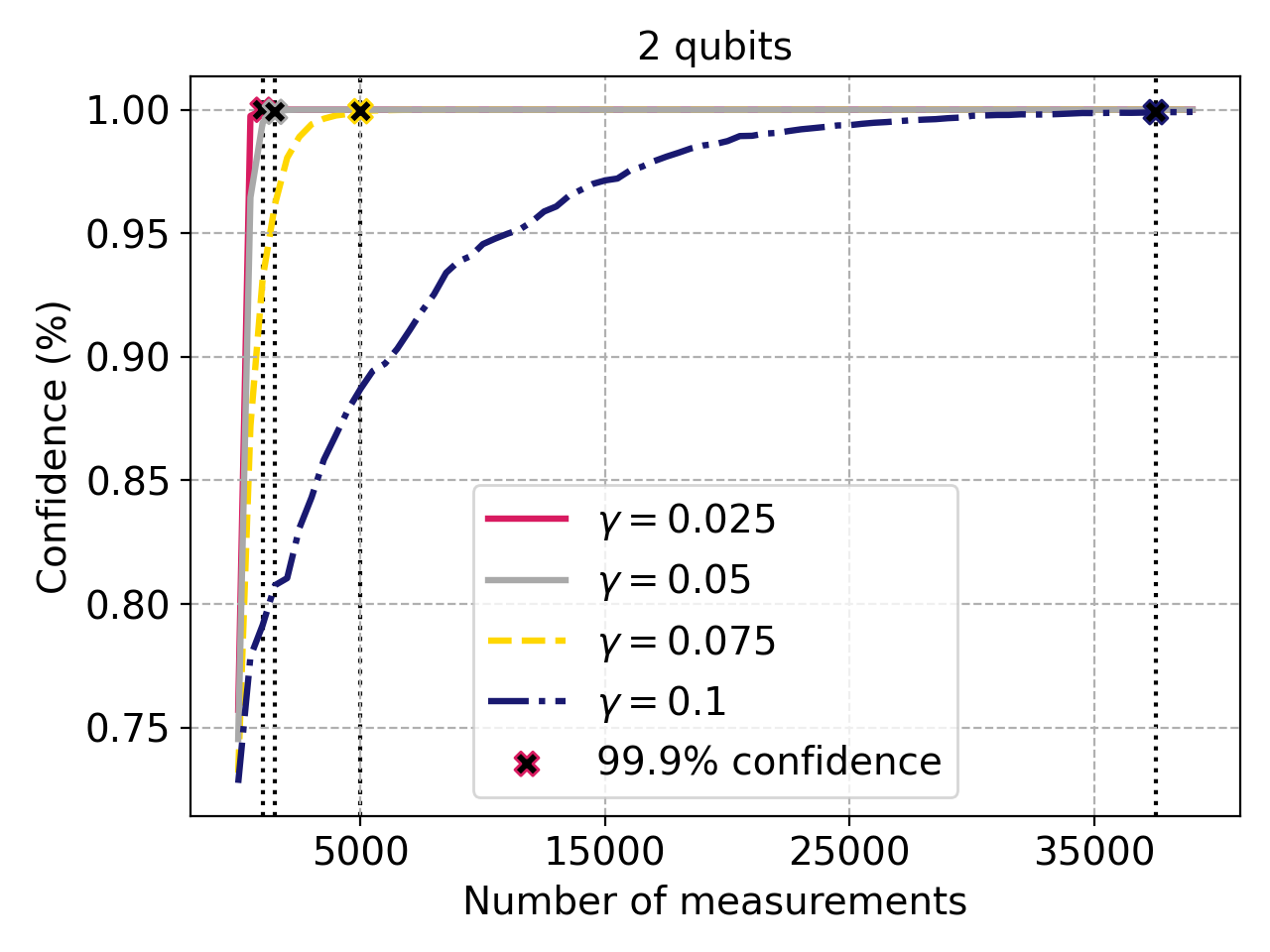}
    % from plot_test_data_n2_gammas
    \caption{Experimental simulation of the QGEM experiment for the 2-qubit parallel setup for different decoherence rates at $\tau=2.5 \,\si{\s}$. With $\gamma = [0.025, 0.050, 0.075, 0.1] \, \si{\hertz}$, respectively, having the witness expectation values $-0.108$, $-0.074$, $-0.043$, $-0.016$ (see figure \ref{fig:decoherencerates}). The number of measurements needed to confirm the entanglement increases for higher decoherence rates.}
    \label{fig:data_n2_gamma}
\end{figure}

\begin{figure}[h!]
    \centering
    \includegraphics[width=\linewidth]{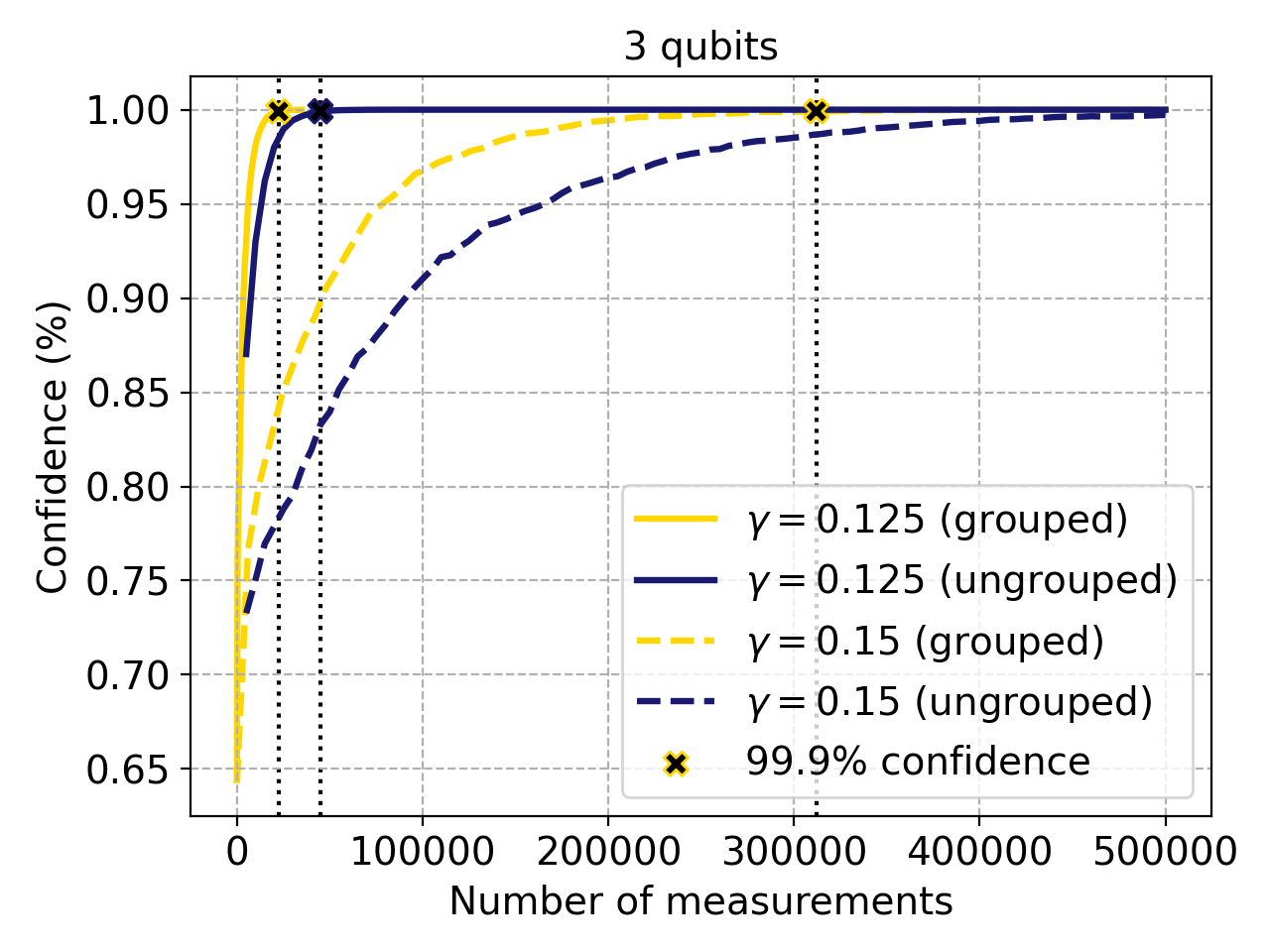}
    \caption{Experimental simulation of the QGEM experiment for the 3-qubit parallel setup at $\tau = 2.5 \, \si{\s}$. For $\gamma = 0.125, 0.15 \, \si{\hertz}$ the witness expectation values are respectively $-0.030$ and $-0.008$. %. For $\gamma = 0.15 \, \si{\hertz}$ the witness expectation value is $-0.008$. 
    The 2-qubit setup cannot detect entanglement for these decoherence rates. Measurements with grouped operators reduce the number of measurements needed to confirm the entanglement.}
    \label{fig:data_n3_gamma_high}
\end{figure}

\begin{table}[t]
\vspace{5mm}
\centering
\begin{center}
\setlength{\tabcolsep}{10pt} % Default value: 6pt
\renewcommand{\arraystretch}{1.5}
\begin{tabular}{||c c c||} 
 \hline
 $\gamma \, (\si{\hertz})$ & $n=2$ & $n=3$\\ [0.5ex] 
 \hline\hline
 0.025 & $\sim 1000$ & $\sim 1500$ \\ 
 \hline
 0.05 & $\sim 1500$ & $\sim 3000$ \\
 \hline
 0.075 & $\sim 5000$ & $\sim 5500$ \\
 \hline
 0.1 & $\sim 37\,500$ & $\sim 13\,000$ \\ 
 \hline
 0.125 & - & $\sim 42\,500$ \\
 \hline
 0.15 & - & $\sim 750\,000$ \\ [1ex]
 \hline
\end{tabular}
\end{center}
\caption{Comparison of the approximate number of measurements needed to confirm the entanglement with $99.9\%$ confidence for the $n=2$ and $n=3$ parallel setups under different decoherence rates at $\tau=2.5 \,\si{\s}$. 
The witness expectations values for $n=2$ for $\gamma = [0.025, 0.050, 0.075, 0.1] \, \si{\hertz}$ are respectively $-0.108$, $-0.074$, $-0.043$, $-0.016$, for $\gamma>0.1 \, \si{\hertz}$ the witness becomes positive.
The witness expectation values for $n=3$ for $\gamma = [0.025$, $0.050$, $0.075$, $0.1$, $0.125$, $0.15] \, \si{\hertz}$ are respectively $-0.156$, $-0.117$, $-0.084$, $-0.055, -0.030, -0.008$.}
\label{table:measurements}
\end{table}

In figure \ref{fig:data_n3_gamma}, the confidence levels for the 3-qubit parallel setup are shown for different decoherence rates $\gamma \in [0.025 - 0.1] \, \si{\hertz}$. 
Comparing this to the 2-qubit setup plotted in figure \ref{fig:data_n2_gamma} we see that the number of measurements needed to confirm  the entanglement with $99.9\%$ confidence is lower for the 2-qubit setup than for the 3-qubit setup for the decoherence rates $0.025 - 0.075 \, \si{\hertz}$, although the difference is not very large.
At higher decoherence rates ($\gamma \geq 0.1 \, \si{\hertz}$) the 3-qubit setup needs fewer measurements compared to the 2-qubit setup.

From figure \ref{fig:decoherencerates} we can predict that the 2-qubit setup has a negative witness at $\tau = 2.5 \, \si{\s}$ for approximately $\gamma<0.12 \, \si{\hertz}$.
For $0.12 \, \si{\hertz} < \gamma < 0.15 \, \si{\hertz}$, we have to look at the 3-qubit setup for measurements. 
The confidence level as a function of the number of measurements for the higher decoherence rates is plotted in figure \ref{fig:data_n3_gamma_high}.

Around $\gamma>0.15 \, \si{\hertz}$, the 3-qubit setup will probably start experiencing too much decoherence for it to detect any entanglement, this can already be seen from the $\gamma=0.15 \, \si{\hertz}$ graph in figure \ref{fig:data_n3_gamma_high}.
The above results are summarised in table \ref{table:measurements}.

\begin{table}[t]
\centering
\begin{center}
\setlength{\tabcolsep}{10pt} % Default value: 6pt
\renewcommand{\arraystretch}{1.5}
\begin{tabular}{||c c c||} 
 \hline

 $\gamma \,(\si{\hertz})$ & $n=2$ & $n=3$\\ [0.5ex] 
 \hline\hline
 0.025 & $\sim 580$ & $\sim 640$ \\ 
 \hline
 0.05 & $\sim 1250$ & $\sim 1390$ \\
 \hline
 0.075 & $\sim 3900$ & $\sim 2900$ \\
 \hline
 0.1 & $\sim 29\,200$ & $\sim 6500$ \\ 
 \hline
 0.125 & - & $\sim 22\,400$ \\
 \hline
 0.15 & - & $\sim 312\,000$ \\ [1ex]
 \hline
\end{tabular}
\end{center}
\caption{Comparison of the approximate number of measurements needed to confirm entanglement with $99.9\%$ confidence for the $n=2$ and $n=3$ parallel setups under different decoherence rates at $\tau=2.5 \,\si{\s}$. The number of measurements needed is reduced by doing the experimental simulation with the grouped operators (see table \ref{table:op_grouping2}), as was discussed in section \ref{sec:witness}.}
\label{table:measurements2}
\end{table}

As discussed before in section \ref{sec:witness}, the number of measurements can be reduced by grouping the operators that need to be measured.
Table \ref{table:op_grouping2} showed the number of Pauli operator groups for different witnesses, an example of the grouping was given in the appendix \ref{appendix:EW}.
By measuring a group of operators simultaneously the total number of measurements can drop a lot, as illustrated in the figure \ref{fig:data_n3_gamma_high} for decoherence rates $\gamma=0.125, 0.15 \, \si{\hertz}$.
The results from the experimental simulation with grouped operators are summarised in the table \ref{table:measurements2}.
Performing measurements yields a marked improvement, especially for the 3-qubit QGEM setup where the number of measurements at lower decoherence rates $\gamma<0.6 \, \si{\hertz}$ become less than or approximately equal to the number of measurements needed for the 2-qubit setup.
This makes the 3-qubit QGEM setup even more favourable.

%%%------------------------------------------------
\section{Qubits vs Qudits}\label{sec:qudits}

In Ref. \cite{tilly2021qudits} qudits (a superposition of $D$ states~\footnote{Experimentally the number of superpositions states $D$ can be seen as the number of arms in each interferometer. The number of particles $n$ are the number of interferometers.}) instead of qubits (which are qudits with $D=2$) were proposed as a way to protect against the decoherence.
Due to the increase in the number of measurements needed to confirm entanglement for qudit setups, they concluded that qudits should be used only if the decoherence is sufficiently high such that the witness for the qubit setup becomes positive ($\gamma>0.12 \, \si{\hertz}$).
In section \ref{sec:measurements} we saw that the three-qubit setup also only outperforms the two-qubit setup for sufficiently high decoherence rates, therefore we compare the two-qudit and three-qubit cases.

\begin{figure}[t]
    \centering
    \includegraphics[width=\linewidth]{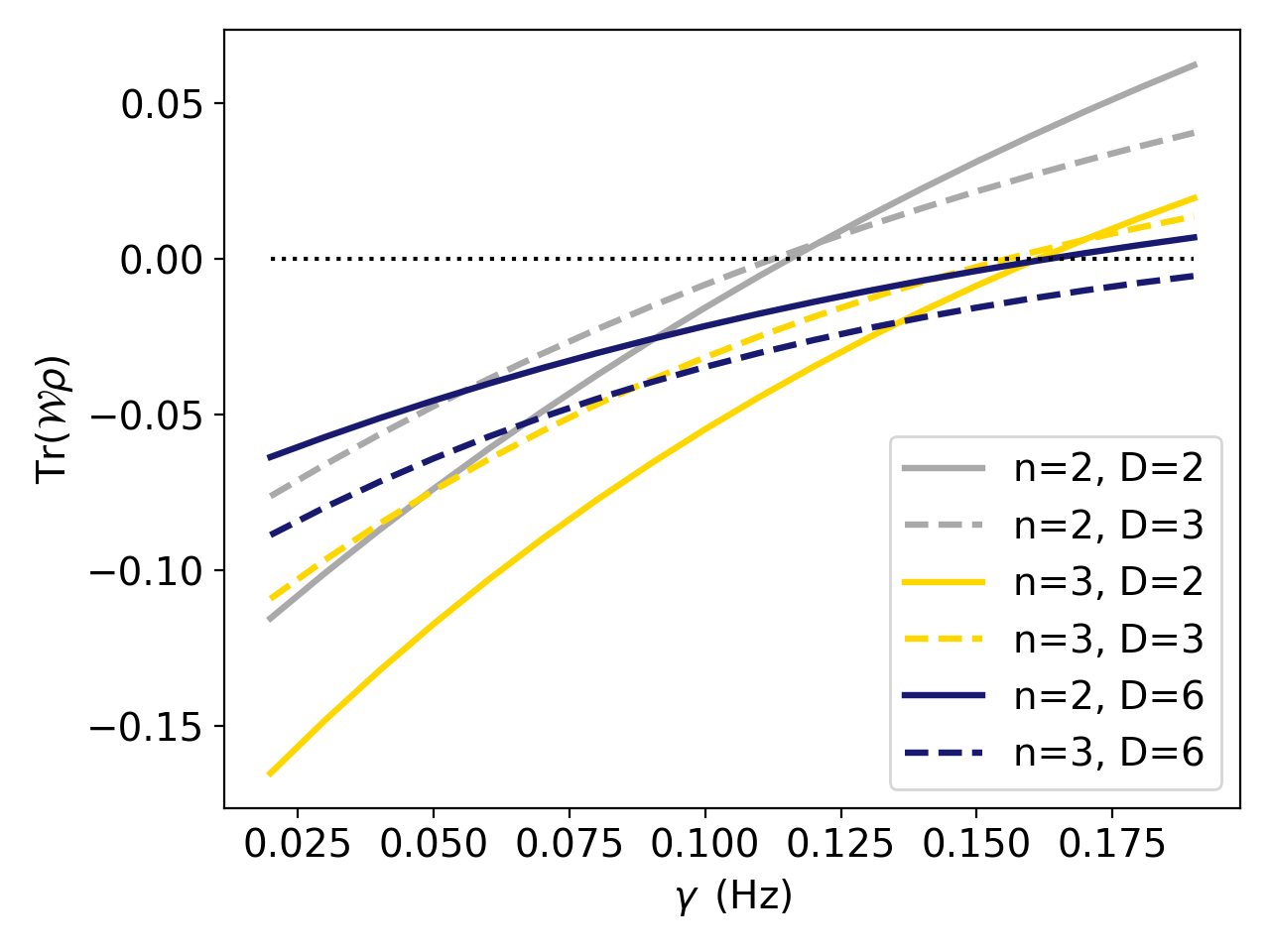} % from new_qgem_model_EW_deco_add.py & some_plots.py
    \caption{The PPT entanglement witness is plotted for different setups at $\tau=2.5 \, \si{\s}$.  
    $n$ is the number of particles 
    and $D$ is the number of superposition states.
    All setups are parallel setups (figure \ref{fig:setup}). 
    The dotted line indicates $\Tr(\mathcal{W}\rho) = 0$.
    Adding a single qubit (increasing $n$ by one) makes the setup more resilient against the decoherence compared to adding a single superposition state (increasing $D$ by one).
    Taking $n=2, D=6$ seems to have the same effect as adding a third qubit ($n=3, D=2$).
    The $n=3, D=6$ setup seems most resilient against decoherence although for this setup the number of needed measurements will be extremely high.}
    \label{fig:EW_qutrits}
\end{figure}

In figure \ref{fig:EW_qutrits}, we compare the entanglement witness expectation value for $n=2$ and $n=3$ qubits ($D=2$) and qutrits (superposition of 3 states, $D=3$). There seems to be very little difference when switching from qubits to qutrits.
It seems that switching from qubits to six-dimensional qudits has the same effect with respect to resilience against the decoherence as switching from the 2-qubit to the 3-qubit setup.
However, the number of operators needed to measure the witness increases much more for the six-dimensional qudit case (94 operator groups) compared to the 3-qubit case (12 operator groups).
The Table \ref{table:op_grouping_qutrits} shows the number of operator (groups) in the witness for the setups considered in the  figure \ref{fig:EW_qutrits}.
From the 2-qubit case adding either one superposition dimension or one qubit does not differ much in the number of operator groups that are added (14 and 12 operator groups respectively), but from figure \ref{fig:EW_qutrits}, we infer that the 3-qubit setup outperforms the 2-qutrit setup.

For a better comparison, we look at the (grouped) measurement plots \ref{fig:data_n3_gamma}, \ref{fig:data_n3_gamma_high}, and compare that with the results from \cite{tilly2021qudits}.
The number of measurements needed to reach a $99.9\%$ confidence level in the entanglement is much higher for the six-dimensional 2-qudit ($n=2, D=6$) setup compared to the 3-qubit setup:
for $\gamma = 0.125 \, \si{\hertz}$, the six-dimensional 2-qudit setup needs about $2,000,000$ measurements (or $200,000$ when grouping operators\footnote{These numbers were derived in \cite{tilly2021qudits}, where a different grouping strategy was used.
It is worth noting that that grouping strategy may not be realisable in the actual experimental setting due to non-local operations.}), while the 3-qubit setup needs about $42,500$ measurements (or $22,400$ when grouping operators).
Ref. \cite{tilly2021qudits} concluded that only for $\gamma>0.12 \, \si{\hertz}$, it is favourable to use qudits over qubits, specifically the six-dimensional qudit setup. However, we found here that the 3-qubit setup is even more favourable for these decoherence rates.
One could also think about switching to 3-qudit setups for improvement, which we will leave for the future investigation.
Looking at the results in \cite{tilly2021qudits}, this would only be necessary for $\gamma>0.16 \, \si{\hertz}$ (when the 3-qubit setup never have a negative witness).
It might also be favourable to add a qubit instead of switching to qudits, this will also be explored in another paper.

\begin{table}[h!]
\centering
\begin{center}
\setlength{\tabcolsep}{10pt} % Default value: 6pt
\renewcommand{\arraystretch}{1.5}
\begin{tabular}{||c c c c||} 
 \hline
 n & D & \#operators & \#operator groups\\ [0.5ex] 
 \hline\hline
 2 & 2 & 4 & 3 \\ 
 \hline
 3 & 2 & 26 & 12 \\
 \hline
 2 & 3 & 77 & 14 \\
 \hline
 2 & 6 & 1272 & 94 \\ [1ex]
 \hline
\end{tabular}
\end{center}
\caption{In this table, we show the number of Pauli operators that make up the decomposed PPT witness, the number of Pauli operator group for the different number of qudits $n$, and different number of superposition states $D$.}
\label{table:op_grouping_qutrits}
\end{table}

%%%------------------------------------------------
\section{Conclusion \& Discussion} \label{sec:conclusion}

For the 3-qubit case we studied three different setups: the parallel, the linear and the star setup (see figure \ref{fig:setup_all}).
We have found that the 3-qubit parallel setup is optimal, it leads to the highest rate of the entanglement generation (figure \ref{fig:entcomp}) and in our chosen basis provides the best witness (figure \ref{fig:witness_over_time}). 
The chosen subsystem matters when analysing the entanglement generation.
Taking the partial trace or partial transpose of the second system is the most favourable since it has a shorter average distance to the other subsystems and therefore the highest interaction and the best generation of the entanglement. 

Sections \ref{sec:witness} and \ref{sec:decoherence} indicated that the 3-qubit setup compared to the 2-qubit setup provides a better witness (i.e. a more negative witness) which is also better resilient to the decoherence (i.e. the witness stays negative for a longer time and at higher decoherence rates). However, the cost of having a better witness by introducing an extra qubit is that it requires more measurements. 
In section \ref{sec:decoherence}, we saw that the number of Pauli operators/operator groups in the witness increases for the 3-qubit case. 
This was reflected in the number of measurements needed (see section \ref{sec:measurements}).
Tables \ref{table:measurements} and \ref{table:measurements2} show that for smaller decoherence rates the 2-qubit setup is favourable in terms of the number of measurements, while for higher decoherence rates the 3-qubit setup is favourable.
The turning point seems to be around $\gamma=0.08 \, \si{\hertz}$, or $\gamma=0.06 \, \si{\hertz}$ when considering grouping the operators.

In appendix \ref{appendix:decoherence_rate}, the decoherence rate was estimated.
It was found that the decoherence rate is expected to be at least  $\gamma=0.05 \, \si{\hertz}$ for environmental temperatures of $T\ge0.5 \, \si{\kelvin}$ (see figure \ref{fig:my_label}).
This is very close to the decoherence rate $\gamma=0.06 \, \si{\hertz}$, for which the 3-qubit setup becomes favourable with respect to the number of measurements needed to confirm the entanglement up to $99.9\%$ confidence level.
Based on the estimation of the decoherence rate, the 3-qubit QGEM protocol provides an improvement in the number of needed measurements compared to the the original QGEM protocol.
However, this is very much dependent on the expected decoherence rate and therefore on the experimental setting.
The decoherence rate increases as the temperature of the environment increases, also it is highly dependent on the size of the superposition particles.

To give a better understanding of how realisable the 3-qubit QGEM protocol is in terms of number of measurements, we convert the number of measurements to experiment time.
The interaction time was taken to be $\tau = 2.5\,\si{\s}$, and the time it takes to construct the superpositions / to bring the superpositions back together is taken to be $0.5 \, \si{\s}$ \cite{bose2017spin}.
For the smallest expected decoherence rate $\gamma = 0.05 \, \si{\hertz}$ $1390$ measurements are needed to confirm entanglement with $99.9\%$ confidence in the 3-qubit parallel setup.
The total experiment time would therefore be approximately $(2.5 \, \si{\s} + 2 * 0.5 \, \si{\s})* 1390 = 4865 \, \si{\s}$ (a similar time is found in the 2-qubit case).
The experiment would take about 1 hour and 20 minutes excluding all the setup that needs to be done between consecutive measurements, this is realisable.
For comparison, at $\gamma = 0.1 \, \si{\hertz}$ the 2-qubit system measurements would take 28.4 hours, while the 3-qubit system measurements take 6.3 hours.
%(and the experiment time at a decoherence rate of $0.15 \, \si{\hertz}$ would be $\sim 3.5 \, \text{days}$).

Additionally, the 3-qubit system was compared to the 2-qudit setups (section \ref{sec:qudits}). 
The 2-qudit setup ($n=2, D=6$) was found to be equally resilient against the decoherence as the 3-qubit setup, but it requires far more measurements.
The gain from switching to qudits is negated by the increase in the number of needed measurements, while for a 3-qubit system the number of measurements stays more feasible. For $\gamma>0.16 \, \si{\hertz}$ the 3-qubit setup cannot witness entanglement anymore, and one could consider switching to a 3-qudit setup. However, since adding a qubit to the setup causes a clear improvement in the rate of entanglement generation, this begs the question of whether a 4-, 5-, or 6-qubit setup could provide further improvements (specifically for $\gamma>0.16 \, \si{\hertz}$)?
We find that indeed increasing the number of qubits increases the system's resilience against the decoherence further.
However, the extra qubits also require more operator groups to be measured and the key is to find a balance between these two to find the optimal experimental system.

%%%------------------------------------------------
\section*{Acknowledgements} \label{sec:acknowledgements}
M.S. is supported by the Fundamentals of the Universe research programme within the University of Groningen.
J.T. is supported by an industrial CASE (iCASE) studentship, funded by and UK EPSRC [EP/R513143/1], in collaboration with University College London and Rahko Ltd.
R.J.M. is supported by a University College London departmental studentship. 
A.M.’s research is funded by the Netherlands Organisation for Science and Research (NWO) grant number 680-91-119. 
SB would like to acknowledge EPSRC grants No. EP/N031105/1 and EP/S000267/1.

%%%------------------------------------------------
\bibliography{bibtex.bib}
\bibliographystyle{ieeetr}

%%%------------------------------------------------

\onecolumngrid
\appendix
%%%------------------------------------------------
\section{Witnessing entanglement} \label{appendix:EW}

For the 2-qubit parallel setup the witness is given by: \cite{chevalier2020witnessing}:
\begin{equation}
    \mathcal{W} = \frac{1}{4} (1 - X \otimes X - Y \otimes Z - Z \otimes Y ).
\end{equation}
In this case, the witness is composed of four operators. The identity term will always have an expectation value of $1$, and therefore does not need to be measured. The three operators left, however, still needed to be measured separately when considering the grouping rules mentioned in the main text. Indeed,  the first Pauli operator in each of the three tensor operators does not commute with one another, therefore failing the condition of qubit-wise commutativity. There are means to still measure these operators together as they generally commute, however, this would require non-local operations \cite{tilly2021qudits}, which we consider not so realistic in an experiment.  

For the 3-qubit parallel setup the witness for the middle qubit is:
\begin{align} \label{eq:EW_three_particles}
    \mathcal{W} &= \frac{1}{26} (1 - 1 \otimes 1 \otimes X - 1 \otimes X \otimes X - 1 \otimes Y \otimes Y 
    - 1 \otimes Y \otimes Z - 1 \otimes Z \otimes Y - X \otimes 1 \otimes 1 - X \otimes X \otimes 1 \nonumber \\
    &- X \otimes X \otimes X - X \otimes Y \otimes Z - X \otimes Z \otimes Y - X \otimes Z \otimes Z  
    - Y \otimes 1 \otimes Y - Y \otimes 1 \otimes Z - Y \otimes X \otimes Y - Y \otimes X \otimes Z \nonumber \\
    &- Y \otimes Y \otimes 1 - Y \otimes Z \otimes 1 - Y \otimes Z \otimes X - Z \otimes 1 \otimes Y 
     - Z \otimes 1 \otimes Z - Z \otimes X \otimes Y - Z \otimes X \otimes Z - Z \otimes Y \otimes 1 \nonumber \\
    & - Z \otimes Y \otimes X - Z \otimes Z \otimes X).
\end{align}
Here we can indeed group the list of $26$ operators ($25$ when discarding the identity term) into groups that commute qubit-wise. As a first example of the rule, consider the first two operators (after the identity) $1 \otimes 1 \otimes X$ and $1 \otimes X \otimes X $.
Each operator in the tensor commutes with the operator of same index in the other tensor: they (qubit-wise) commute, and can clearly be both measured a the same time (if one collects measurements for $1 \otimes X \otimes X $, they can use the measurements on the last operator to infer the expectation value of  $1 \otimes 1 \otimes X$).
Based on this, we can group all the $25$ terms above in $12$ groups presented in the table \ref{tab:grouping_example}.

\begin{table}[ht] 
    \centering
    \setlength{\tabcolsep}{10pt} % Default value: 6pt
    \renewcommand{\arraystretch}{1.5}
    \begin{tabular}{p{0.005\linewidth}p{0.05\linewidth} |c}
    \hline 
    & 1 & $1 \otimes 1 \otimes X \quad 1 \otimes X \otimes X \quad X \otimes 1 \otimes 1 \quad X \otimes X \otimes 1 \quad X \otimes X \otimes X$ \\
    & 2 & $1 \otimes Y \otimes Y \quad Y \otimes 1 \otimes Y \quad Y \otimes Y \otimes 1$ \\ 
    & 3 & $1 \otimes Y \otimes Z \quad X \otimes Y \otimes Z$ \\ 
    & 4 & $1 \otimes Z \otimes Y \quad X \otimes Z \otimes Y$ \\ 
    & 5 & $Y \otimes 1 \otimes Z \quad Y \otimes X \otimes Z$ \\ 
    & 6 & $Y \otimes Z \otimes 1 \quad Y \otimes Z \otimes X$ \\ 
    & 7 & $Z \otimes 1 \otimes Y \quad Z \otimes X \otimes Y$ \\ 
    & 8 & $Z \otimes 1 \otimes Z \quad Z \otimes X \otimes Z$ \\ 
    & 9 & $Z \otimes Y \otimes 1 \quad Z \otimes Y \otimes X$ \\
    & 10 & $X \otimes Z \otimes Z$ \\ 
    & 11 & $Y \otimes X \otimes Y$ \\ 
    & 12 & $Z \otimes Z \otimes X$ \\
    \hline
    \end{tabular}
    \caption{Example of qubit-wise commuting groups constructed using the operators composing the entanglement witness of the three-qubit setup (eq. \ref{eq:EW_three_particles})}
    \label{tab:grouping_example}
\end{table}
%%%------------------------------------------------
\section{Estimating the Decoherence Rate} \label{appendix:decoherence_rate}

In \cite{van2020quantum} an explicit expression for $\gamma$ was derived for the 2-qubit QGEM setup. 
They considered decoherence due to the scattering between the environmental particles and the superpositions.
Since the environment state (the state of the scattered particle) is dependent on the position of the test particle, the scattered particle shares information about the superposition state. 
For a large number of scatterings - so for a large scattering rate and/or after a long time - the system will decohere. 
We will use the result from Ref. \cite{van2020quantum} for the 3-qubit setup discussed in this paper.
The 3 superpositions are assumed to be spatially unaffected by scattering with the environment and independent of each other.

Following \cite{schlosshauer2007decoherence, van2020quantum}, we write the final state of the scattered particle that scattered off the system's position state $\ket{\vec{x}}$ as:
\begin{equation}
    \ket{E(\Vec{x})}
    = e^{- i \Vec{q}\Vec{x} / \hbar} \hat{S}_0 e^{i \Vec{q}\Vec{x} / \hbar} \ket{E_j}
\end{equation}
Here, $\vec{q}$ is the momentum of the scattered environmental particle and $S_0$ is the scattering operator acting on the scattering centre.
The density matrix element of the superposition with the scattered particle(s) traced out is given as:
\begin{equation}
    \rho_S(\vec{x},\vec{x}') 
    = \ket{\vec{x}}\bra{\vec{x}'} \bra{E(\Vec{x})}\ket{E(\Vec{x}')}
\end{equation}
This equation is similar to eq. \eqref{eq:rhodeco2}. 
It is clear that the diagonal terms $\vec{x}=\vec{x}'$ are not affected by the environment.
The expression for $\bra{E(\Vec{x})}\ket{E(\Vec{x}')}$ was evaluated in the S-matrix formalism in \cite{schlosshauer2014quantum}. 

We will consider the decoherence due to the scattering of the air molecules with the superposition, and scattering with and absorption/emission of the blackbody photons. These are thought to be the leading causes of decoherence for a macroscopic spatial superposition \cite{van2020quantum, schlosshauer2014quantum, romero2011quantum}. For these decoherence sources, we can simplify the decoherence rate because for the environmental temperature considered here they have either a much longer or shorter wavelength compared to the superposition width $\Delta x$.

As discussed in sections \ref{sec:setup} and \ref{sec:EE}, we consider $m \sim 10^{-14} \, \si{\kg}$, $\Delta x = 250 \, \si{\mu \m}$ for the setup; furthermore, we take the internal temperature of the system to be $T_i = 0.15 \, \si{\kelvin}$, the pressure of the environment to be $10^{-15} \, \si{\pascal}$, and the test masses to be micro-crystals (diamond) \cite{bose2017spin}.

For an environmental temperature $T_e \sim 0.15 \, \si{\kelvin}$, the wavelength of blackbody photons is $\lambda_{bb} \sim 30 \, \si{\mm}$ \cite{romero2011quantum}.
This is much longer than the superposition width, therefore the blackbody radiation can be evaluated in the long-wavelength limit. 
At the temperature $T_e \sim 0.15 \, \si{\kelvin}$, the wavelength of the air molecules is $\lambda_{\text{air}} \sim 0.82 \, \si{\nano\m}$ (with the mass of the scattering particle taken to be $28.97 \text{u}$, which is the mass of a typical air molecule in the atomic mass units.) \cite{romero2011quantum}.
This is much shorter than the superposition width, and the scattering with air molecules can therefore be evaluated in the short-wavelength limit.

For a constant superposition width, Refs. \cite{van2020quantum, schlosshauer2007decoherence, schlosshauer2014quantum} found the decoherence rate due to short-wavelength and long-wavelength particles to be:
\begin{equation}
    \gamma = 
    \Gamma_{\text{air}} + \Lambda_\text{bb} \Delta x^2 \, ,
\end{equation}
with $\Gamma_{\text{air}}$ the total scattering rate with the air molecules (short-wavelength limit contribution). $\Lambda_\text{bb}$ is a scattering constant dependent on the density, velocity and effective cross-section for the blackbody radiation (long-wavelength limit contribution).
The scattering rate $\Gamma_{\text{air}}$ depends also on the average velocity of the air molecules, the radius of the superposition qubit (taken to be of micron size with radius a = $10^{-6} \, \si{m}$), and the number density of the particles (taken to be $N/V = 10^8 \, \si{\m^{-3}}$, which corresponds to the vacuum pressure $P=10^{-15}\,\si{\pascal}$).
For our setup, we find
\begin{equation}
    \Gamma_\text{air}
    \sim 0.03 \, \si{\hertz} \, .
\end{equation}
Where we used that $\Gamma_\text{air} =  \lambda_\text{air}^2  \Lambda_\text{air}$ \cite{romero2011quantum} and that $\Lambda_\text{air} \sim 4\times 10^{16} \, \si{\s^{-1} \m^{-2}}$ for the pressure, test mass size and temperature mentioned previously.
We assumed that the environment is a perfect gas, and used the formula provided in \cite{schlosshauer2007decoherence}:
\begin{equation}\label{eq:gamma_est1}
    \Lambda_{\text{air}} 
    = \frac{8}{3\hbar^2} \frac{N}{V} a^2 \sqrt{2 \pi m_\text{air}} (k_B T_e)^{3/2}. %(3.73)
\end{equation}
Now consider the scattering of the system with the blackbody photons. We find that
\begin{equation}\label{eq:}
    \Lambda_{\text{bb}} 
    \sim 0.003 \, \si{\s^{-1}\m^{-2}} \, .
\end{equation}
This consists of the blackbody scattering $\Lambda_s$, emission $\Lambda_e$, and absorption $\Lambda_a$, $\Lambda_{\text{bb}}  = \Lambda_s + \Lambda_e + \Lambda_a$.
The scattering constants are determined by using the results found in \cite{schlosshauer2007decoherence}, which models the scattering constant for the blackbody radiation and a particle modelled as the dielectric spheres with the radii $a$, and the dielectric constant $\epsilon$.
It is dependent also on Boltzmann's constant $k_b$, the reduced Planck's constant $\hbar$, and the speed of light $c$.
\begin{equation}\label{eq:lambda_s}
    \Lambda_s = 8! \frac{8}{9\pi} a^6 c \, \Re \left(\frac{\epsilon-1}{\epsilon+2} \right)^2 \left( \frac{k_b T_e}{\hbar c}\right)^9 \zeta(9)
\end{equation}
The scattering constant depends heavily on the size of the test mass and the temperature of the environment.
The constants were taken as described previously, furthermore $\zeta(n)$ is the Riemann $\zeta$-function, and the dielectric constant for diamond is $\epsilon = 5.68+1.1\times10^{-4} i$.
This gives
\begin{equation}
    \Lambda_s
    \sim 3\times 10^{-8} \, \si{\s^{-1}\m^{-2}} \, .
\end{equation}
The scattering constants $\Lambda_e$ and $\Lambda_a$ were determined in \cite{van2020quantum} by adjusting eq. \eqref{eq:lambda_s} to use the probability of emission/absorption instead of the cross section of scattering. These are given by:
\begin{equation}\label{eq:gamma_est2}
    \Lambda_{e(a)} = \frac{16 \pi^5}{189} a^3 c \, \Im \left(\frac{\epsilon-1}{\epsilon+2} \right) \left( \frac{k_b T_{i(e)}}{\hbar c}\right)^6 \, ,
\end{equation}
note that for $\Lambda_e$ we should use the temperature of the test mass ($T_i=0.15 \, \si{\kelvin}$), while for $\Lambda_a$ we should use the temperature of the environment (also set to $T_e=0.15 \, \si{\kelvin}$).
We find that
\begin{equation}
    \Lambda_e =
    \Lambda_a
    \sim 0.003 \, \si{\s^{-1}\m^{-2}}.
\end{equation}
Adding all the blackbody decoherence sources together we get the value in eq. \eqref{eq:}.
Due to the blackbody photons, we thus have $\Gamma_{\text{bb}}  = \Lambda_{\text{bb}}  \Delta x^2 \sim 4\times 10^{-10} \, \si{\hertz}$.

\begin{figure}[t]
    \centering
    \begin{subfigure}[t]{0.49\textwidth}
    \includegraphics[width=\textwidth]{deco_estimation.png}
    \caption{}
    \end{subfigure}
    \hfill
    \begin{subfigure}[t]{0.49\textwidth}
    \includegraphics[width=\textwidth]{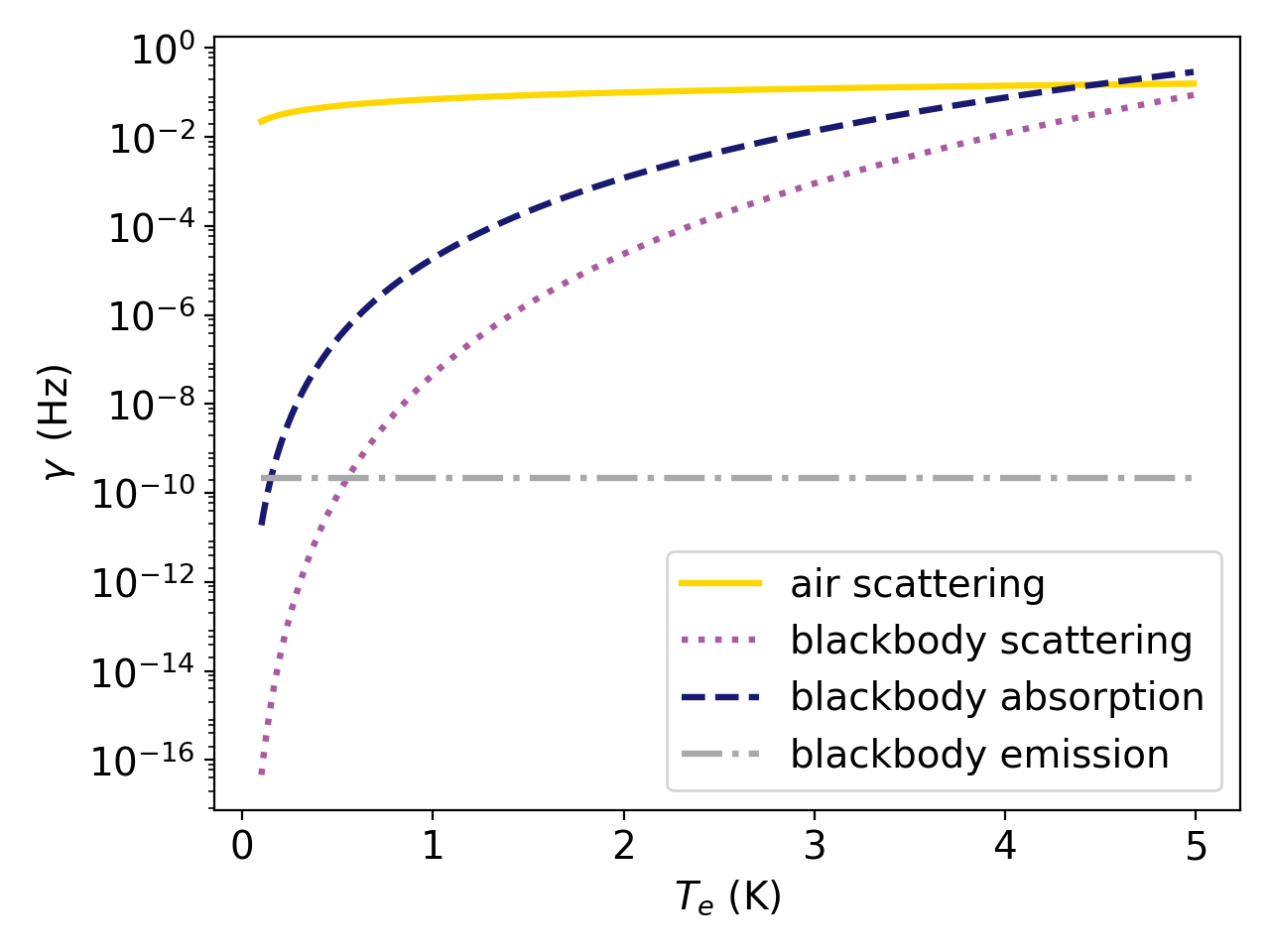}
    \caption{}
    \end{subfigure}
    \caption{An estimate of the decoherence rate due to interaction with the air molecules and the blackbody photons. For environmental temperatures $T_e = 0.1 \, \si{\kelvin} - 5 \, \si{\kelvin}$ the wavelength of the air molecules is much smaller ($1-15 \, \si{\nano\m}$) and the wavelength of the blackbody photons is much larger ($3-98 \, \si{\mm}$) than the superposition width ($250 \, \si{\micro \m}$).
    Therefore, we can use respectively the short- and long-wavelength limits to study their decoherence effects for this range of temperatures.
    For the decoherence rate due to the scattering off the air molecules the number density for the gas is taken to be $N/V = 10^8 \, \si{\m^{-3}}$. 
    The radius of the particles is taken to be $a = 10^{-6} \, \si{\m}$, and we used the dielectric constant for diamond, with the internal temperature $T_i=0.15 \, \si{\kelvin}$.
    %The left plot is also shown in section \ref{sec:decoherence}, the right plot shows the same data but with a logarithmic y-axis [{\bf I do not see this in the plot] to emphasise that the contribution due to the blackbody emission is nonzero.
    (a) This plot was used in section \ref{sec:decoherence}, it is based on the calculations performed in this appendix.
    (b) The same data is plotted as in the figure to the left but with a logarithmic scale on the y-axis.
    From this figure we see clearly that the contribution to the decoherence rate from the blackbody emission is nonzero.  
    It has a constant value of $\sim 10^{-10} \, \si{\hertz}$ since it is independent of the external temperature $T_e$.
    }
    \label{fig:my_label}
\end{figure}

For a higher environmental temperature of $T_e = 1 \, \si{\kelvin}$, the approximations of the short- and long-wavelength limits still hold, and we find
\begin{align*}
    \Lambda_e &\sim 0.003 \, \si{\s^{-1}\m^{-2}}
    &\Lambda_a \sim 0.3 \times 10^{3} \, \si{\s^{-1}\m^{-2}} \\
    \Lambda_s &\sim 0.7 \, \si{\s^{-1}\m^{-2}} 
    &\Lambda_{\text{air}} \sim 7\times 10^{17} \, \si{\s^{-1}\m^{-2}}
\end{align*}
so that $\Gamma_{\text{bb}}  \sim 10^{-5} \, \si{\hertz}$ and $\Gamma_{\text{air}}  \sim 0.07 \, \si{\hertz}$.
The suspected decoherence rate is dependent on the temperature of the environment.
In figure \ref{fig:my_label} on the next page the decoherence rate is plotted as a function of the environmental temperature.

Figure \ref{fig:my_label} depends greatly on the assumed experimental parameters.
The expected decoherence rate decreases (increases) when the number density $N/V$, the size of the superposition $a$, and the external temperature $T_e$ are decreased (increased).
Changing the material (and thus the dielectric constant $\epsilon$) also changes the expected decoherence rate, although it only influences the decoherence rate due to the blackbody photons, which for $T \leq 3 \, \si{\kelvin}$ is dominated by the decoherence due to the scattering with air molecules.
As can be seen from eqs. \eqref{eq:gamma_est1}, \eqref{eq:lambda_s} and \eqref{eq:gamma_est2}, the decoherence rate is highly sensitive to the size of the test particle and the temperature of the environment.

So far we have considered decoherence sources during the part of the experiment where the qubits are interacting for a time $\tau_\text{int}$. 
Additionally, decoherence plays a role while creating and bringing back together the superpositions. 
In this paper, the decoherence during the creating/reuniting of the superpositions has not been taken into account. 
If we were to take this into account, we would also have to include the entanglement generation during these steps. 
The influence of the decoherence and the entanglement generation during creation and destruction of the superposition should be the subject of a separate paper.

\newpage
%%%------------------------------------------------
\section{Supplementary Analysis}\label{appendix:EE}

\begin{figure}[h]
    \centering
    \begin{minipage}{0.48\textwidth}
        \centering
        \includegraphics[width=\linewidth]{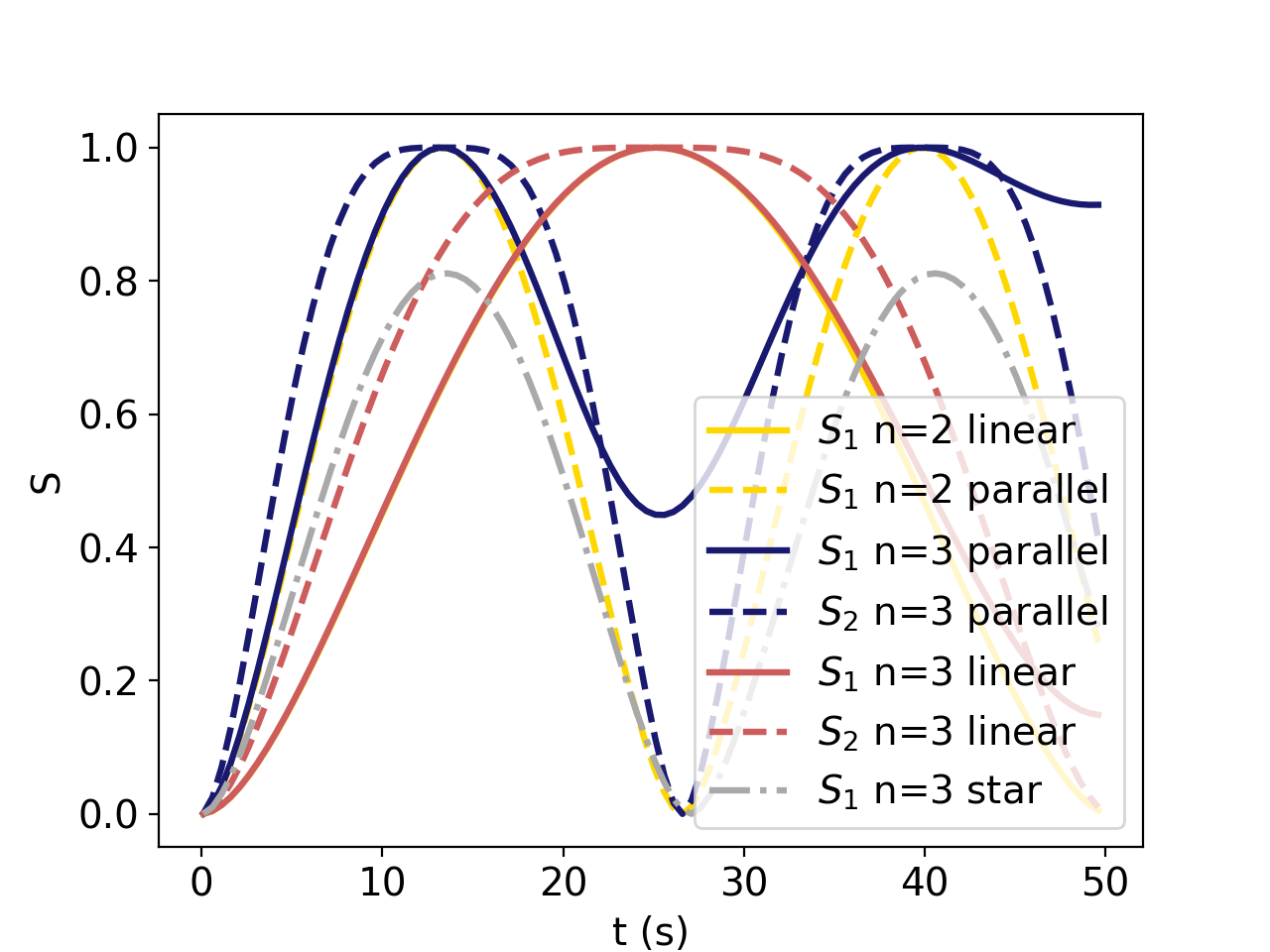}
        % from entanglement2.py
        \caption{Comparison of the entanglement entropy generated within $\tau\in[0,50]\,\si{\s}$. Note that due to symmetry for the parallel and linear setup we have $S_1 = S_3$ and for the star setup we have $S_1 = S_2 = S_3$. 
        The lines representing $S_1$ $n=3$ linear and $S_1$ $n=2$ linear almost overlap.
        During the first $10 \, \si{\s}$ the rate of entanglement generated by the $S_2$ $n=3$ parallel setup is highest, however this time scale is unrealistic experimentally.}
        \label{fig:entcomp_large_t}
    \end{minipage}\hfill
    \begin{minipage}{0.48\textwidth}
        \centering
        \includegraphics[width=\linewidth]{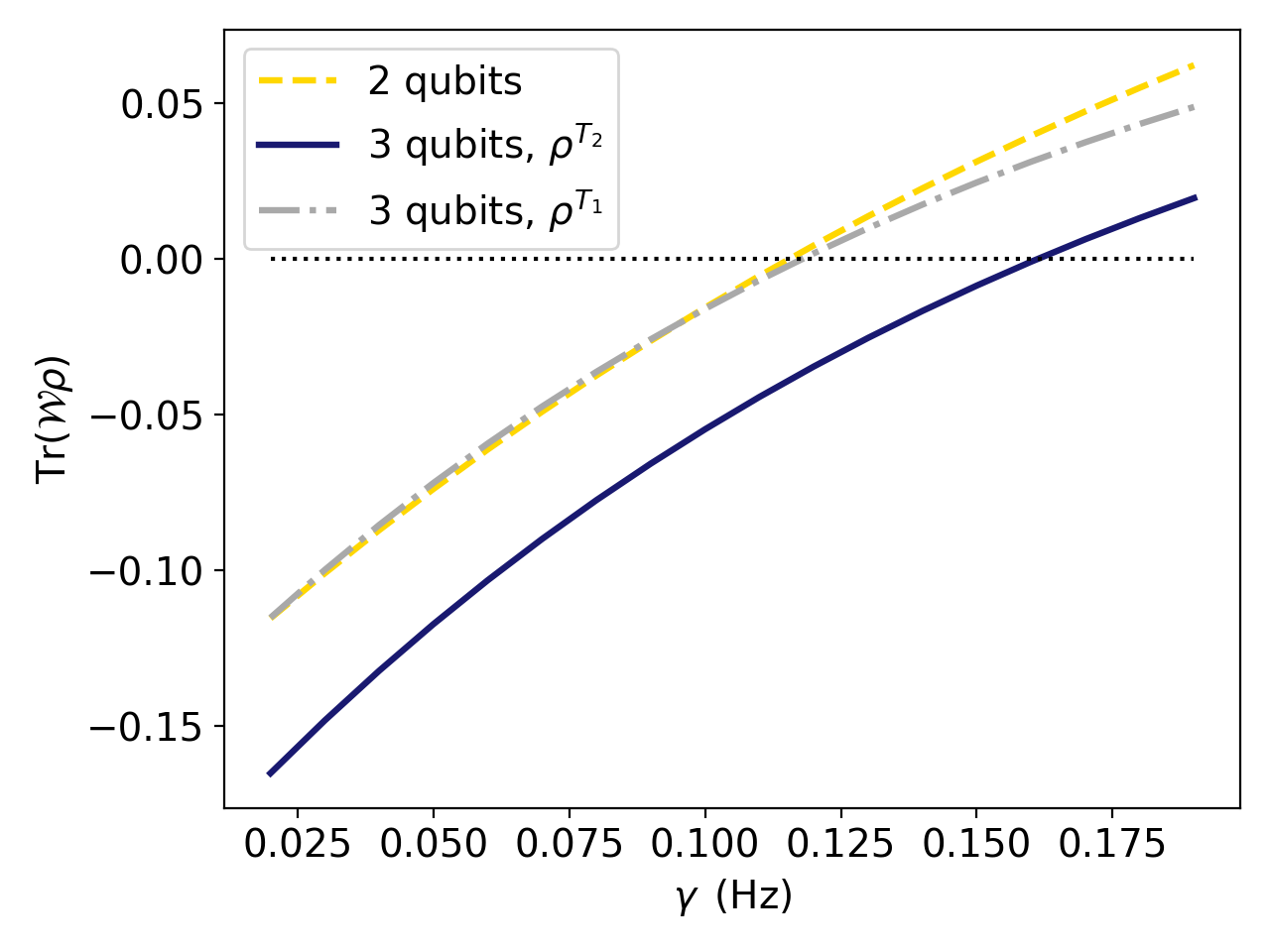}
        \caption{The PPT entanglement witness expectation value is plotted for the 2- and 3-qubit parallel setups, for $\gamma \in [0.02,0.20] \, \si{\hertz}$ and at $\tau=2.5 \, \si{\s}$. The partial transpose for the 3-qubit setup can be taken over the second subsystem (denoted $\rho^{T_2}$) or over the first subsystem (denoted $\rho^{T_1}$, which due to the symmetry of the setup is the same as $\rho^{T_3}$).
        The partition $\rho^{T_2}$ is negative for higher decoherence rates compared to the 2-qubit setup and $\rho^{T_1}$.
        Taking the partial transpose over the first subsystem does not give an improvement in the witness.
        The dotted line indicates $\Tr(\mathcal{W}\rho) = 0$.}
        \label{fig:extra_EW}
    \end{minipage}
\end{figure}

In figure \ref{fig:entcomp} the entanglement entropy was shown as a function of time for $\tau\in[0,5] \, \si{\s}$.
For an extended period of time, we find the result given in figure \ref{fig:entcomp_large_t}. This time frame is of course not realistic for an experiment, however in the reader's interest we wanted to illustrate the cyclical behaviour of entanglement entropy.  \newline

Figure \ref{fig:extra_EW} is supplemental to the analysis performed in section \ref{sec:EE} which showed that taking the partial trace over the outer systems provided the highest rate of entanglement generation.
From figure \ref{fig:entcomp} we saw that the rate of entanglement generation for the three-qubit system $S_1$ compared to the two-qubit system almost overlap, while the rate of entanglement generation for the three-qubit system $S_2$ is clearly higher.
Figure \ref{fig:extra_EW} below shows the same but in terms of the witness.
The witness expectation value of the three-qubit setup with the partial transpose of the first subsystem $\rho^{T_1}$ almost overlaps with the witness expectation value for the two-qubit setup.
Taking the partial transpose of the middle subsystem $\rho^{T_2}$ provides a witness that is more resistant against decoherence.

%\begin{figure}[h!]
%    \centering
%    \includegraphics[width=0.43\linewidth]{Figure_larget2.png}
%    \caption{Comparison of the entanglement entropy generated within $\tau\in[0,50]\,\si{\s}$. Note that due to symmetry for the parallel and linear setup we have $S_1 = S_3$ and for the star setup we have $S_1 = S_2 = S_3$.  The lines representing $S_1$ $n=3$ linear and $S_1$ $n=2$ linear almost overlap. During the first $10 \, \si{\s}$ the rate of entanglement generated by the $S_2$ $n=3$ parallel setup is highest, however this time scale is unrealistic experimentally.}
%    \label{fig:entcomp_large_t}
%\end{figure}
%
%\begin{figure}[h!]
%    \centering
%    \includegraphics[width=0.43\linewidth]{EW_gamma_n23_3.png}
%    \caption{The PPT entanglement witness expectation value is plotted for the 2- and 3-qubit parallel setups, for $\gamma \in [0.02,0.20] \, \si{\hertz}$ and at $\tau=2.5 \, \si{\s}$. The partial transpose for the 3-qubit setup can be taken over the second subsystem (denoted $\rho^{T_2}$) or over the first subsystem (denoted $\rho^{T_1}$, which due to the symmetry of the setup is the same as $\rho^{T_3}$). The partition $\rho^{T_2}$ is negative for higher decoherence rates compared to the 2-qubit setup and $\rho^{T_1}$. Taking the partial transpose over the first subsystem does not give an improvement in the witness. The dotted line indicates $\Tr(\mathcal{W}\rho) = 0$.}
%\label{fig:extra_EW}
%\end{figure}
%%%------------------------------------------------

\end{document}